\RequirePackage{fix-cm}
\documentclass[twocolumn, epjc3, comma, sort&compress, natbib]{svjour3}
\bibpunct{[}{]}{,}{n}{}{,} 

\journalname{Eur. Phys. J. C}

\usepackage{latexsym}
\usepackage{amsmath}
\usepackage{amssymb}
\usepackage{amsfonts}
\usepackage{CJKutf8}

\usepackage[mathscr,scaled=1.15]{urwchancal}
\DeclareFontFamily{OT1}{pzc}{}
\DeclareFontShape{OT1}{pzc}{m}{it}%
{<-> s * [1.15] pzcmi7t}{}
\DeclareMathAlphabet{\mathpzc}{OT1}{pzc}{m}{it}

\usepackage{color}

\usepackage{supertabular}
\usepackage{placeins}
\usepackage{epsfig}
\usepackage{graphicx}

\definecolor{purple}{rgb}{0.5,0,0.5}
\definecolor{blue}{rgb}{0.0,0,0.9}
\definecolor{prdblue}{rgb}{0.133,0.118,0.498}
\usepackage[colorlinks=true, pdfstartview=FitV, linkcolor=prdblue, citecolor= prdblue, urlcolor=prdblue]{hyperref}

\begin{document}
\begin{CJK*}{UTF8}{gbsn}

\title{$\,$\\[-6ex]\hspace*{\fill}{\normalsize{\sf\emph{Preprint no}.\
NJU-INP 099/25}}\\[1ex]
Kaon and Pion Fragmentation Functions}

\author{Hui-Yu Xing (邢惠瑜)\thanksref{NJU,INP}%
    $\,^{\href{https://orcid.org/0000-0002-0719-7526}{\textcolor[rgb]{0.00,1.00,0.00}{\sf ID}}}$
\and
    Wen-Hao Bian (边文浩)\thanksref{NJU,INP}%
       $\,^{\href{https://orcid.org/0009-0005-9980-3376}{\textcolor[rgb]{0.00,1.00,0.00}{\sf ID}},}$
\and
    \\Zhu-Fang Cui (崔著钫)\thanksref{NJU,INP}%
       $\,^{\href{https://orcid.org/0000-0003-3890-0242}{\textcolor[rgb]{0.00,1.00,0.00}{\sf ID}},}$
\and
    Craig D.\ Roberts\thanksref{NJU,INP}%
       $\,^{\href{https://orcid.org/0000-0002-2937-1361}{\textcolor[rgb]{0.00,1.00,0.00}{\sf ID}},}$
}

\institute{School of Physics, \href{https://ror.org/01rxvg760}{Nanjing University}, Nanjing, Jiangsu 210093, China \label{NJU}
\and
Institute for Nonperturbative Physics, \href{https://ror.org/01rxvg760}{Nanjing University}, Nanjing, Jiangsu 210093, China
\label{INP}
%
           %
\\[1ex]
Email:
\href{mailto:phycui@nju.edu.cn}{phycui@nju.edu.cn} (ZFC);
\href{mailto:cdroberts@nju.edu.cn}{cdroberts@nju.edu.cn} (CDR)
            }

\date{2025 September 02} 

\maketitle

\end{CJK*}

\begin{abstract}
The Drell-Levy-Yan relation is employed to obtain pion and kaon elementary fragmentation functions (EFFs) from the hadron-scale parton distribution functions (DFs) of these mesons.  Two different DF sets are used: that calculated using a symmetry-preserving treatment of a vector\,$\times$\,vec\-tor contact interaction (SCI) and the other expressing results obtained using continuum Schwinger function methods (CSMs).  Thus determined, the EFFs serve as driving terms in a coupled set of hadron cascade equations, whose solution yields the complete array of hadron-scale fragmentation functions (FFs) for pion and kaon production in high energy reactions.  After evolution to scales typical of experiments, the SCI and CSM FF predictions are seen to be in semiquantitative agreement.  Importantly, they conform with a range of physical expectations for FF behaviour on the endpoint domains $z\simeq 0, 1$, \emph{e.g}., nonsinglet FFs vanish at $z=0$ and singlet FFs diverge faster than $1/z$.  Predictions for hadron multiplicities in jets are also delivered.  They reveal SU$(3)$ symmetry breaking in the char\-ged-kaon/neutral-kaon multiplicity ratio, whose size diminishes with increasing reaction energy, and show that, with increasing energy, the pion/kaon ratio in $e^+ e^- \to h X$ diminishes to a value that is independent of hadron masses.
 \end{abstract}

\section{Introduction}
Jets of energetic hadrons are often produced in high energy interactions.  The particles in such a jet have nearly parallel longitudinal momenta and relatively small transverse momenta.  Within quantum chromodynamics \linebreak (QCD), they are understood to be created by gluon and quark partons, which, after being produced in the initial collision, escape the interaction region and, under the influence of confinement dynamics, fragment into a shower of colourless hadrons \cite{Field:1976ve, Field:1977fa, Altarelli:1981ax, Ellis:1991qj, Metz:2016swz, Chen:2023kqw}.  The process of parton$\,\to\,$hadron ($p\to h$) conversion -- hadronisation -- is described by fragmentation functions (FFs), which may be interpreted as probability densities.   For instance, $D_q^{h}\!(z;\zeta) dz$ is the probability that, in an interaction characterised by an energy scale $\zeta$, a $q$ quark escaping the collision region produces a hadron $h$, giving up a light-front fraction $z$ of its pre-emission momentum.
A common reinterpretation sees $D_q^{h}\!(z;\zeta) dz$ as the number of $h$ hadrons inside the $q$ quark within the momentum fraction range $[z,z+dz]$ at the scale $\zeta$.
Ideally, FFs are universal, \emph{i.e}., independent of the type of collision that produces the partons.
The following conditions are sufficient for this to be true: each elementary $p\to h$ fragmentation function is entirely determined by the wave function of $h$ and each emission in a cascade is independent of its predecessor.

Assuming the validity of various factorisation theorems \cite{Ellis:1991qj},
FF models are usually built via phenomenological analyses of selected hadron production data -- see, \emph{e.g}., Refs.\,\cite{Hirai:2007cx, deFlorian:2014xna, Bertone:2017tyb, Soleymaninia:2020bsq, Moffat:2021dji, AbdulKhalek:2022laj, Gao:2024dbv}.  However, existing inferences have large uncertainties.  This is a problem because FFs appear in the convolution formulae for many cross-sections that are used to infer parton distribution functions\linebreak (DFs); hence, precise knowledge will be necessary if optimal use is to be made of new data obtained at existing and anticipated accelerator facilities \cite{Aguilar:2019teb, BESIII:2020nme, Anderle:2021wcy, Arrington:2021biu, Schnell:2022nlf, Quintans:2022utc}.  Consequently, both the need for and importance of reliable theoretical FF predictions are magnified.

Like DFs, however, FFs are essentially nonperturbative objects.  Hitherto, few realistic calculations have been available.  Owing to their innate timelike character, the numerical simulation of lattice-regularised QCD (lQCD) is ill-suited to FF computation.  A path to their calculation is provided by continuum Schwinger function methods (CSMs)
\cite{Roberts:2021nhw, Binosi:2022djx, Ding:2022ows, Ferreira:2023fva, Raya:2024ejx}; and a contemporary treatment of pion FFs is given in Ref.\,\cite{Xing:2023pms}.

It is worth stressing that whilst confinement is often mentioned when discussing FFs, the associated\linebreak meaning is usually not made explicit.  Instead, a loose link between FFs and confinement dynamics is drawn implicitly via reference to the transitions from coloured to colour-singlet objects, which are involved in the had\-ro\-nisation process.  Part of the problem is that, even today, an agreed practicable definition of confinement is lacking -- see, \emph{e.g}., Ref.\,\cite[Sect.\,5]{Ding:2022ows}.
Notwithstanding that, it should be possible to draw tighter connections through the calculation of FFs using CSMs, which enable the exploration of various confinement scenarios.
In particular, CSMs link confinement with emergent hadron mass (EHM) phenomena
\cite{Roberts:2021nhw, Binosi:2022djx, Ding:2022ows, Ferreira:2023fva, Raya:2024ejx, Salme:2022eoy, Krein:2023azg}, whose elucidation is a goal of an array of experimental programmes \cite{Aguilar:2019teb, BESIII:2020nme, Anderle:2021wcy, Arrington:2021biu, Schnell:2022nlf, Quintans:2022utc, Carman:2023zke, Mokeev:2023zhq}.

Herein, we extend the approach of Ref.\,\cite{Xing:2023pms} to the simultaneous prediction of both pion and kaon FFs.  The foundations for these calculations are provided by crossing symmetry and the Drell-Levy-Yan (DLY) relation \cite{Drell:1969jm, Drell:1969wd, Drell:1969wb, Gribov:1972ri, Gribov:1972rt}, which together enable one to obtain hadron-scale, $\zeta=\zeta_{\cal H}$, elementary $q\to h$ FFs, $d_{\mathpzc q}^h$, from $q$-in-$h$ DFs, ${\mathpzc q}^h$, \textit{viz}.\
\begin{equation}
    d_{\mathpzc q}^h(z;\zeta_{\cal H}) = z {\mathpzc q}^h(1/z ;\zeta_{\cal H})\,.
    \label{DLYR}
\end{equation}
These elementary FFs (EFFs) are then used in cascade equations to obtain the complete FFs
\cite{Field:1976ve, Field:1977fa}.
Calculated in this way, the FFs are universal because the procedure satisfies the sufficiency conditions stated above.

In addition, it is important to note that Eq.\,\eqref{DLYR} means all manifestations of EHM in ${\mathpzc q}^h$ are also expressed in the source function which drives $q\to h$ fragmentation.  This information flows into the full fragmentation function via the hadron cascade equations.  Thus, not unexpectedly, perhaps, the seeds of confinement, as expressed in hadronisation, can already be found in the wave functions of the hadrons involved.

Our discussion is arranged as follows.
A symmetry-preserving treatment of a vector\,$\times$\,vec\-tor contact interaction (SCI) \cite{Gutierrez-Guerrero:2010waf} is used in Sect.\,\ref{SecEFFs} to establish a range of EFF concepts and results.
Hadron jet cascade equations for pion ($\pi$) and kaon ($K$) production are introduced and discussed in Sect.\,\ref{HJEs}.  Empirically, fragmentation to pions and kaons is almost exhaustive.
Section~\ref{FFEvolution} explains the all-orders (AO) approach to FF scale evolution.  It also discusses the momentum sum rule and how inclusion of a gluon FF ensures that sum rule is obeyed.
Potential sources of systematic uncertainty in our ultimate FF predictions are discussed and quantified in Sect.\,\ref{SysUnc}.
Solutions of the SCI cascade equations are described in Sect.\,\ref{SecSCIFFs}, which also demonstrates explicitly that the momentum sum rules are satisfied.
Section~\ref{SecCSMFFs} explains how realistic EFFs are obtained from CSM predictions for $\pi, K$ DFs, describes solutions to the cascade equations defined therewith, and compares the CSM predictions with some contemporary phenomenological fits.
Predictions for $\pi, K$ relative and individual multiplicities in $e^+ e^- \to h X$ reactions are discussed in Sect.\,\ref{HadronMultiplicities} and compared with available data.
Section~\ref{Epilogue} presents a summary and perspective.

\section{Elementary Fragmentation Functions: SCI}
\label{SecEFFs}
%
%
To begin, it is worth presenting the EFFs obtained using a SCI \cite{Gutierrez-Guerrero:2010waf}.  In the chiral limit, \emph{i.e}., when the quark current masses are zero, one obtains the following hadron scale valence quark DF \cite{Zhang:2020ecj}:
${\mathpzc u}^{\pi^+}(x;\zeta_{\cal H}) = 1$;
and, via Eq.\,\eqref{DLYR}:
\begin{equation}
\label{SCIEFF}
d_{\mathpzc u}^{\pi^+}(z;\zeta_{\cal H}) = 2 z\,.
\end{equation}
There is unit probability that the parton generates a hadron; so, the EFF is normalised such that
\begin{equation}
\int_0^1\,dz\, d_{\mathpzc q}^{h}(z;\zeta_{\cal H})=1\,.
\label{dnorm}
\end{equation}

At the zeroth stage of any cascade, a $u$ quark can produce both $\pi^{+}, \pi^{0}$:
\begin{subequations}
\begin{align}
D_{{\mathpzc u}0}^{\pi^+}(z;\zeta_{\cal H}) & =
d_{\mathpzc u}^{\pi^+}(z;\zeta_{\cal H})\,, \\
D_{{\mathpzc u}0}^{\pi^0}(z;\zeta_{\cal H}) & =
\tfrac{1}{2} d_{\mathpzc u}^{\pi^+}(z;\zeta_{\cal H})\,.
\end{align}
\end{subequations}
The different weighting owes to isospin.

{\allowdisplaybreaks
Generalising to nonzero quark current masses -- we assume isospin symmetry throughout, the SCI yields the following expression for a $u$-in-$h=\pi^+, K^+$ DF, $q=d,s$, $M_{q_\pm}=M_q\pm M_u$:
\begin{equation}
	\label{pidfexp}
	{\mathpzc u}_{h^+}(x ; \zeta_H)=\frac{N_c}{4 \pi^2}(c_{EE} E_h^2 +c_{EF} E_h F_h + c_{FF} F_h^2),
\end{equation}
with
\begin{subequations}
\label{SCIuh}
	\begin{align}
		c_{EE} & =\bar{\mathpzc C}_1\left(\varsigma_0\right) +x(1-x)\left[m_h^2-M_{q_-}^2\right]\frac{2 \bar{\mathpzc C}_2\left(\varsigma_0\right)}{\varsigma_0}\,,	\\
	c_{EF}&=-\frac{M_{q_+}\left[x M_u+(1-x) M_q\right]}{M_q M_u}\bar{\mathpzc C}_1\left(\varsigma_0\right) \nonumber \\
    & \qquad -\frac{ x(1-x)M_{q_+}^2\left[m_h^2-M_{q_-}^2\right]}{M_s M_u}\frac{2 \bar{\mathpzc C}_2\left(\varsigma_0\right)}{\varsigma_0}\,,	\\
c_{FF}&=\frac{(1-2 x)M_{q_+}^3 M_{q_-}}{4 M_q^2 M_u^2}\bar{\mathpzc C}_1\left(\varsigma_0\right) \nonumber \\
& \qquad +\frac{ x(1-x)M_{q_+}^4\left[m_h^2-M_{q_-}^2\right]}{4 M_q^2 M_u^2} \frac{2 \bar{\mathpzc C}_2\left(\varsigma_0\right)}{\varsigma_0}\,,
	\end{align}
\end{subequations}	
where
$\varsigma_0=x M_q^2+(1-x)M_u^2-x(1-x)m_h^2$;
$M_{u,q}$ are dressed-quark masses, obtained from the SCI gap equation;
$E_h$, $F_h$ are constants that specify the SCI bound-state amplitude of the $h$-meson, obtained by solving the SCI Bethe-Salpeter equation; and ($n\in {\mathbb Z}^\geq$)
\begin{align}
n !\, \overline{\cal C}^{\rm iu}_n(\sigma) & = \Gamma(n-1,\sigma \tau_{\textrm{uv}}^{2}) - \Gamma(n-1,\sigma \tau_{\textrm{ir}}^{2})\,,
\label{eq:Cn}
\end{align}
where $\Gamma(\alpha,y)$ is the incomplete gamma function.

Owing to isospin symmetry and the nature of the hadron scale \cite{Cui:2020tdf}:
\begin{equation}
\label{sKdfexp}
	{\mathpzc s}_{\bar K^0}(x ; \zeta_H) = {\mathpzc s}_{K^-}(x ; \zeta_H) = {\mathpzc u}_{K^+}(1-x ; \zeta_H)\,;
\end{equation}
and, by charge conjugation,
$\bar {\mathpzc s}_{K^+}(x ; \zeta_H)
= \bar {\mathpzc s}_{K^0}(x ; \zeta_H)
= {\mathpzc s}_{K^-}(x ; \zeta_H)$.
}

\begin{table}[t]
\caption{\label{Tab:DressedQuarks}
SCI couplings, $\alpha_{\rm IR}/\pi$, ultraviolet cutoffs, $\Lambda_{\rm uv}$, and current-quark masses, $m_q$, $q=u/d,s$, that deliver a good description of $\pi$, $K$ pseudoscalar meson properties, along with the dressed-quark masses, $M_q$, meson masses, $m_{P}$, and leptonic decay constants, $f_{P}$, they produce; all obtained with $m_G=0.5\,$GeV, $\Lambda_{\rm ir} = 0.24\,$GeV when defining the SCI.
The calculated Bethe-Salpeter amplitude coefficient functions are:
$E_\pi=3.59$, $F_\pi=0.47$;
$E_K=3.70$, $F_K=0.55$.
Empirically, at a sensible level of precision \cite{ParticleDataGroup:2024cfk}:
$m_\pi =0.14$, $f_\pi=0.092$;
$m_K=0.50$, $f_K=0.11$.
%
(We assume isospin symmetry and list dimensioned quantities in GeV.  Details are available in Ref.\,\cite{Xing:2022sor}.)}
\begin{center}
\begin{tabular*}
{\hsize}
{
l@{\extracolsep{0ptplus1fil}}|
c@{\extracolsep{0ptplus1fil}}|
c@{\extracolsep{0ptplus1fil}}
c@{\extracolsep{0ptplus1fil}}
c@{\extracolsep{0ptplus1fil}}
|c@{\extracolsep{0ptplus1fil}}
c@{\extracolsep{0ptplus1fil}}
c@{\extracolsep{0ptplus1fil}}}\hline
& quark & $\alpha_{\rm IR}/\pi\ $ & $\Lambda_{\rm uv}$ & $m$ &   $M$ &  $m_{P}$ & $f_{P}$ \\\hline
$\pi\ $  & $l=u/d\ $  & $0.36\phantom{2}$ & $0.91\ $ & $0.0068\ $ & 0.37$\ $ & 0.14 & 0.10  \\\hline
$K\ $ & $\bar s$  & $0.33\phantom{2}$ & $0.94\ $ & $0.16\phantom{77}\ $ & 0.53$\ $ & 0.50 & 0.11 \\\hline
\end{tabular*}
\end{center}
\end{table}

\begin{figure}[t]
\leftline{\large\sf A}
\vspace*{-2ex}

\centerline{%
\includegraphics[clip, width=0.95\linewidth]{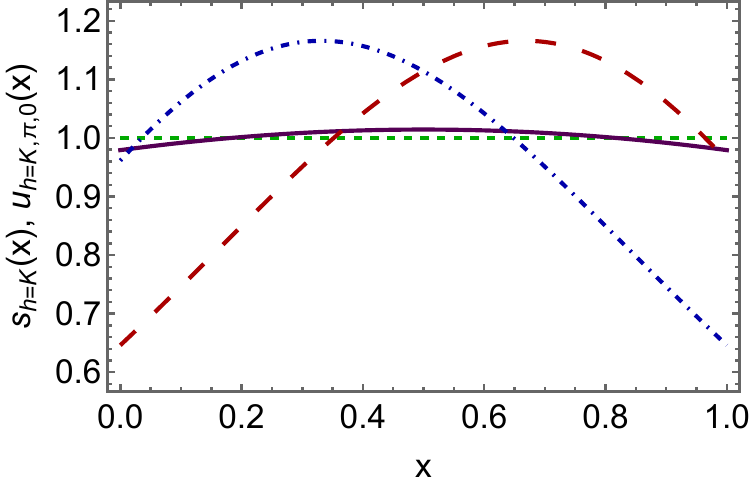}}

\vspace*{1ex}

\leftline{\large\sf B}
\vspace*{-2ex}

\centerline{%
\hspace*{-1ex}\includegraphics[clip, width=0.96\linewidth]{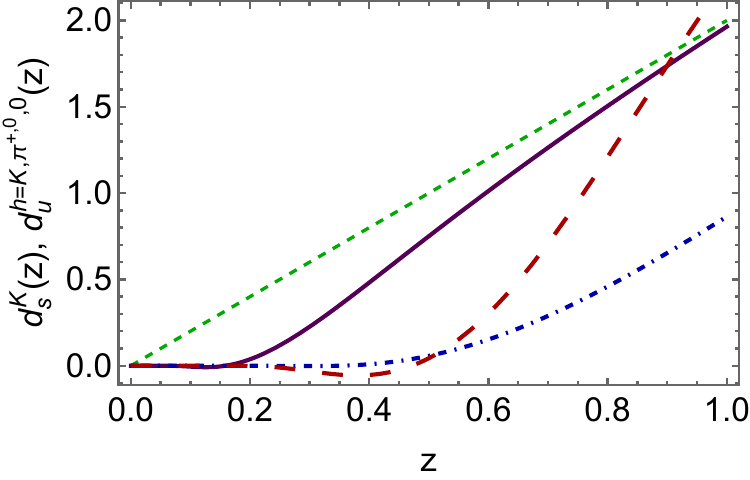}}
\caption{\label{SCIpiKDFs}
{\sf Panel A}.
SCI valence quark parton distribution functions, obtained using Eqs.\,\eqref{pidfexp}, \eqref{SCIuh}, \eqref{sKdfexp}, and the results listed in Table~\ref{Tab:DressedQuarks}:
${\mathpzc s}_{K^-}(x ; \zeta_H)$ -- long-dashed red curve;
${\mathpzc u}_{K^+}(x ; \zeta_H)$ -- dot-dashed blue ;
${\mathpzc u}_{\pi^+}(x ; \zeta_H)$ -- solid purple;
${\mathpzc u}_{\pi^+}(x ; \zeta_H)$ in chiral limit  ($h=0$) -- dashed green.
{\sf Panel B}.
SCI elementary fragmentation functions, obtained from the results in {\sf Panel A} using Eq.\,\eqref{DLYR}.
$d_{\mathpzc s}^{K^-}(x ; \zeta_H)$ -- long-dashed red curve;
$d_{\mathpzc u}^{K^+}(x ; \zeta_H)$ -- dot-dashed blue ;
$d_{\mathpzc u}^{\pi^++\pi^0}(x ; \zeta_H)$ -- solid purple;
$d_{\mathpzc u}^{\pi}(x ; \zeta_H)$ in chiral limit -- dashed green.
%
}
\end{figure}

Recent SCI applications, including details of various calculations, can be found in Refs.\,\cite{Xing:2022sor, Cheng:2022jxe, Yu:2024qsd, Cheng:2024gyv}.
Profiting from those studies, in Table~\ref{Tab:DressedQuarks}, we list each quantity in Eqs.\,\eqref{pidfexp}, \eqref{SCIuh} that is relevant for both the pion and kaon.
Using these values, one obtains the $\pi$ and $K$ valence quark DFs drawn in Fig.\,\ref{SCIpiKDFs}\,A.
Owing to the momentum-independence of the SCI, the hadron scale DFs do not vanish at the endpoints $x\simeq 0, 1$.  Insofar as the illustrations herein are concerned, this artefact is largely immaterial.  It is eliminated by using an interaction that becomes weaker with increasing momentum transfer \cite{Lu:2021sgg}, such as that which underlies the realistic DFs we also consider herein \cite{Cui:2020tdf}.


Using Eq.\,\eqref{DLYR} and the DFs in Fig.\,\ref{SCIpiKDFs}\,A, one obtains the EFFs drawn in Fig.\,\ref{SCIpiKDFs}\,B.
Since a $u$ quark can directly produce $\pi^{+}, \pi^{0}$ and $K^+$, then, generalising Eq.\,\eqref{dnorm}, the associated elementary EFFs are normalised as follows:
\begin{align}
\int_0^1 dz \, \left[ \tfrac{3}{2} d_{\mathpzc u}^{\pi^+}(z;\zeta_{\cal H}) + d_{\mathpzc u}^{K^+}(z;\zeta_{\cal H})\right] = 1\,.
\label{NormEFF}
\end{align}
The associated SCI elementary $u$ quark multiplicities are:
{\allowdisplaybreaks
\begin{subequations}
\label{SCIEFFmq}
\begin{align}
m_u^\pi & = \int_0^1 dz\, \tfrac{3}{2} d_{\mathpzc u}^{\pi^+}(z;\zeta_{\cal H}) = 0.80 \,, \\
m_u^K & = \int_0^1 dz\, d_{\mathpzc u}^{K^+}(z;\zeta_{\cal H}) = 0.20 \,.
\end{align}
\end{subequations}
On the other hand, the $s$ quark can produce $K^-,\bar K^0$; so,
\begin{align}
\int_0^1 dz \, 2 d_{\mathpzc s}^{K^-}(z;\zeta_{\cal H})  = 1\,.
\label{NormsKEFF}
\end{align}
}

\section{Hadron Jet Equations}
\label{HJEs}
We follow Ref.\,\cite{Field:1977fa} in building complete FFs from EFFs.  Namely, with the EFF describing the first fragmentation event for parton $p$ generating hadron $h$ with momentum fraction $z$, then the complete FF, $D_{\mathpzc p}^h(z)$, is obtained via a recursion relation that resums the exhaustive series of such events:
\begin{equation}
D_{\mathpzc q}^h(z) =
d_{\mathpzc q}^h(z) + \sum_{{\mathpzc q}^\prime = {\mathpzc u}, {\mathpzc d}, {\mathpzc s}}\int_z^1 (dy/y)
d_{\mathpzc q}^{{\mathpzc q}^\prime}(z/y) D_{{\mathpzc q}^\prime}^h(z)\,,
\label{JetEq}
\end{equation}
where $h=\pi^\pm, \pi^0, K^\pm, K^0, \bar K^0$.
In all these equations, as explained in Ref.\,\cite{Xing:2023pms}, the resolving scale $\zeta=\zeta_{\cal H}$.

It is worth highlighting some features of the solutions to Eq.\,\eqref{JetEq}.
First,
\begin{equation}
\label{largez}
D_{\mathpzc p}^h(z) \stackrel{z\simeq 1}{\approx} d_{\mathpzc p}^h(z)
\end{equation}
because if the parton gives all its momentum to $h$, then there is none left to contribute to a cascade.
Moreover, one may readily establish algebraically that, for a given parton species, $p$,
\begin{equation}
    \sum_h \int_0^1 dz\,z\, D_p^h(z) = 1\,,
    \label{MomSumRule}
\end{equation}
where the sum runs over all hadrons contained in the shower.  This identity merely states that the hadron jet generated by the parton $p$ contains all the momentum of that initial state, neither more nor less.  Finally \cite{Field:1976ve}:
\begin{equation}
D_{\mathpzc p}^h(z) \stackrel{z \simeq 0}{=} \frac{1}{z}\,,
\label{FeynmanIR}
\end{equation}
because it costs nothing to produce hadrons with zero fraction of the initial parton momentum.  In practice, the impact of this infrared divergence is tamed by had\-ron masses.

Working in the ${\mathpzc G}$-parity symmetry limit \cite{Lee:1956sw}:
{\allowdisplaybreaks
\begin{subequations}
\label{dqqp}
\begin{align}
d_{\mathpzc u}^{{\mathpzc u}}(z) & = d_{\mathpzc u}^{\pi^0}(1-z)
= d_{\mathpzc d}^{{\mathpzc d}}(z)
= d_{\bar{\mathpzc u}}^{\bar{\mathpzc u}}(z)
= d_{\bar{\mathpzc d}}^{\bar{\mathpzc d}}(z)\,, \\
d_{\mathpzc u}^{{\mathpzc d}}(z) & = d_{\mathpzc u}^{\pi^+}(1-z)
= d_{\mathpzc d}^{{\mathpzc u}}(z)
= d_{\bar{\mathpzc u}}^{\bar{\mathpzc d}}(z)
= d_{\bar{\mathpzc d}}^{\bar{\mathpzc u}}(z)\,, \\
d_{\mathpzc u}^{{\mathpzc s}}(z) & = d_{\mathpzc u}^{K^+}(1-z)
= d_{\mathpzc d}^{{\mathpzc s}}(z)
= d_{\bar{\mathpzc u}}^{\bar{\mathpzc s}}(z)
= d_{\bar{\mathpzc d}}^{\bar{\mathpzc s}}(z)\,, \\
d_{\mathpzc s}^{{\mathpzc u}}(z) & = d_{\mathpzc s}^{K^-}(1-z)
= d_{\mathpzc s}^{\bar K^0}(1-z)
= d_{\mathpzc s}^{{\mathpzc d}}(z)\,.
\end{align}
\end{subequations}
Further capitalising on ${\mathpzc G}$-parity symmetry and temporarily ignoring gluon and heavier quark degrees-of-freedom, then the complete FFs must satisfy the following identities:
\begin{subequations}
\label{GParity}
\begin{align}
D_{\mathpzc u}^{\pi^+}(z) & = D_{\bar{\mathpzc d}}^{\pi^+}(z)
= D_{\bar{\mathpzc u}}^{\pi^-}(z) = D_{{\mathpzc d}}^{\pi^-}(z) \,,\\
D_{{\mathpzc u}}^{K^+}(z) & = D_{{\mathpzc d}}^{K^0}(z)
= D_{\bar{\mathpzc u}}^{K^-}(z)  = D_{\bar{\mathpzc d}}^{\bar K^0}(z)\,,\\
D_{{\mathpzc s}}^{K^-}(z) & = D_{{\mathpzc s}}^{\bar K^0}(z)
= D_{\bar{\mathpzc s}}^{K^+}(z) = D_{\bar{\mathpzc s}}^{K^0}(z) \\
D_{{\mathpzc u}}^{\pi^-}(z) &
= D_{{\mathpzc d}}^{\pi^+}(z)
= D_{\bar{\mathpzc u}}^{\pi^+}(z)
= D_{\bar {\mathpzc d}}^{\pi^-}(z)
\,, \\
D_{{\mathpzc u}}^{K^0}(z) & = D_{{\mathpzc d}}^{K^+}(z)
= D_{\bar{\mathpzc u}}^{\bar K^0}(z)  = D_{\bar{\mathpzc d}}^{K^-}(z)\,,\\
D_{{\mathpzc u}}^{K^-}(z) & = D_{{\mathpzc d}}^{\bar K^0}(z)
= D_{\bar{\mathpzc u}}^{K^+}(z)  = D_{\bar{\mathpzc d}}^{K^0}(z)\,, \\
D_{{\mathpzc u}}^{\bar K^0}(z) & = D_{{\mathpzc d}}^{K^-}(z)
= D_{\bar{\mathpzc u}}^{K^0}(z)  = D_{\bar{\mathpzc d}}^{K^+}(z)\,,\\
D_{s}^{{K}^0}(z) &  = D_{{s}}^{K^+}(z) = D_{\bar{s}}^{\bar K^0}(z)  = D_{\bar s}^{K^-}(z) \,,
\label{sKpm00}\\
D_{{\mathpzc u}}^{\pi^0}(z) & = D_{{\mathpzc d}}^{\pi^0}(z)
= D_{\bar{\mathpzc u}}^{\pi^0}(z)  = D_{\bar{\mathpzc d}}^{\pi^0}(z)\,.
\end{align}
\end{subequations}
The first three rows describe the cases in which the hadronising quark or antiquark can be a valence degree-of-freedom in the produced hadron (favoured);
the next five rows, those situations when it cannot (unfavoured);
and the final row, when any initial quark or antiquark fla\-vour can be a valence part of the emitted pion (neutral).
}

Exploiting these identities, Eq.\,\eqref{JetEq} expands to a system of nine coupled equations:
{\allowdisplaybreaks
\begin{subequations}
\label{JetExplicit}
\begin{align}
D_{\mathpzc u}^{\pi^+}(z) & = d_{\mathpzc u}^{\pi^+}(z)
+ \int_z^1\frac{dy}{y}
\sum_{q=u,d,s}\!\!\!  d_{\mathpzc u}^{\mathpzc q}(\frac{z}{y})  D_{\mathpzc q}^{\pi^+}(y)\,,\\
D_{\mathpzc u}^{K^+}\!(z) & = d_{\mathpzc u}^{K^+}(z)
+ \int_z^1\frac{dy}{y}
\sum_{q=u,d,s} \!\!\! d_{\mathpzc u}^{\mathpzc q}(\frac{z}{y})  D_{\mathpzc q}^{K^+}\!(y)\,,\\
D_{\mathpzc s}^{K^-}(z) & = d_{\mathpzc s}^{K^-}(z)
+ \int_z^1\frac{dy}{y}
\sum_{q=u,d}\!\!\!  d_{\mathpzc s}^{\mathpzc q}(\frac{z}{y})  D_{\mathpzc q}^{K^-}(y)\,,
\label{FFstoKm}\\
D_{\mathpzc u}^{\pi^-}(z) & = 0
+ \int_z^1\frac{dy}{y}
\sum_{q=u,d,s} \!\!\!  d_{\mathpzc u}^{\mathpzc q}(\frac{z}{y})  D_{\mathpzc q}^{\pi^-}(y)\,,\\
D_{\mathpzc u}^{K^-}(z) & = 0
+ \int_z^1\frac{dy}{y}
\sum_{q=u,d,s} \!\!\!  d_{\mathpzc u}^{\mathpzc q}(\frac{z}{y})  D_{\mathpzc q}^{K^-}(y)\,,\\
D_{\mathpzc u}^{K^0}(z) & = 0
+ \int_z^1\frac{dy}{y}
\sum_{q=u,d,s} \!\!\!  d_{\mathpzc u}^{\mathpzc q}(\frac{z}{y})  D_{\mathpzc q}^{K^0 }(y)\,,\\
D_{\mathpzc u}^{\bar K^0}(z) & = 0
+ \int_z^1\frac{dy}{y}
\sum_{q=u,d,s} \!\!\!  d_{\mathpzc u}^{\mathpzc q}(\frac{z}{y})  D_{\mathpzc q}^{\bar K^0 }(y)\,,\\
D_{\mathpzc s}^{\pi^+}(z) & = 0
+ \int_z^1\frac{dy}{y}
\sum_{q=u,d}\!\!\!  d_{\mathpzc s}^{\mathpzc q}(\frac{z}{y})  D_{\mathpzc q}^{\pi^+}(y)\,,
\label{FFstopi}\\
D_{\mathpzc s}^{K^+}(z) & = 0
+ \int_z^1\frac{dy}{y}
\sum_{q=u,d}\!\!\!  d_{\mathpzc s}^{\mathpzc q}(\frac{z}{y})  D_{\mathpzc q}^{K^+}(y)\,.
\end{align}
\end{subequations}
}

In order to be used in analysing data, one must employ evolution equations \cite[DGLAP]{Altarelli:1981ax}
to map the hadron scale FFs to some $\zeta > m_p$ ($m_p$ is the proton mass), whereat various factorisation theorems are valid.  In this process, one works with the following singlet ($S$) and nonsinglet ($N$) combinations:
{\allowdisplaybreaks
\begin{subequations}
\label{EvolutionCombinations}
\begin{align}
D_{S_{\mathpzc q}}^\pi(z) & = \frac{3}{2} \left[ D_{\mathpzc q}^{\pi^+}(z) + D_{\bar{\mathpzc q}}^{\pi^+}(z)\right]\,, \label{Spi} \\
D_{N_{\mathpzc q}}^\pi(z) & = \frac{3}{2} \left[ D_{\mathpzc q}^{\pi^+}(z) - D_{\bar{\mathpzc q}}^{\pi^+}(z)\right] \,, \label{Npi}\\
D_{S_{\mathpzc q}}^{K^+}(z) & =  D_{\mathpzc q}^{K^+}(z) + D_{\bar{\mathpzc q}}^{K^+}(z)\,, \label{SK}\\
D_{N_{{\mathpzc q}\neq s}}^{K^+}(z) & =   D_{\mathpzc q}^{K^+}(z) - D_{\bar{\mathpzc q}}^{K^+}(z)\,,
\label{NKq} \\
D_{N_{{\mathpzc s}}}^{K^+}(z) & =   D_{\bar {\mathpzc s}}^{K^+}(z) - D_{{\mathpzc s}}^{K^+}(z)\,,
\label{NKs}
\end{align}
\end{subequations}
As above, the $3/2$ is an isospin Clebsch-Gordon factor.  In order to reconstruct all FFs in the ${\cal G}$-parity symmetry limit, Eq.\,\eqref{GParity}, one needs $q=u,s$ for the pion and $q=u,d,s$ for the kaon.
}

Combining the DLY relation, Eq.\,\eqref{DLYR}, with Eq.\,\eqref{largez}, one learns that the $z\simeq 1$ behaviour of hadron-scale FFs is the same as that of the associated valence quark DF on $x\simeq 1$.
In QCD, this means \cite{Brodsky:1994kg, Yuan:2003fs, Holt:2010vj, Cui:2021mom, Cui:2022bxn, Lu:2022cjx}: $D^h(z;\zeta_{\cal H}) \propto (1-z)^2$.
Since the large-$z$ power increases under evolution, then any QCD-consistent favoured FF should behave as follows:
\begin{equation}
\label{FFzone}
D_{{S_q},{N_q}}^h(z;\zeta) \stackrel{z\simeq 1}{\propto} (1-z)^{2+\gamma(\zeta)},
\end{equation}
where $\gamma(\zeta>\zeta_{\cal H}) \geq 0$ grows logarithmically with $\zeta$.  The powers on glue and sea FFs are, respectively, one and two units greater \cite{Brodsky:1994kg, Yuan:2003fs, Holt:2010vj, Cui:2021mom, Cui:2022bxn, Lu:2022cjx}.
As with analyses of data that attempt to infer DFs, however, these constraints are typically overlooked in phenomenological FF extractions.

\section{Fragmentation Function Evolution}
\label{FFEvolution}
\subsection{AO scheme}
We evolve FFs according to the scheme discussed in Ref.\,\cite[Sec.\,6]{Xing:2023pms}, which adapts the AO approach to DF evolution that is explained in Ref.\,\cite{Yin:2023dbw}.  The AO scheme extends DGLAP evolution \cite{Dokshitzer:1977sg, Gribov:1971zn, Lipatov:1974qm, Altarelli:1977zs} onto QCD's nonperturbative domain.  It has proven efficacious, with an array of successful applications, \emph{e.g}.,
delivering unified predictions for all pion, kaon, and proton DFs \cite{Cui:2020tdf, Chang:2022jri, Lu:2022cjx, Cheng:2023kmt, Yu:2024qsd},
a tenable species separation of nucleon gravitational form factors \cite{Yao:2024ixu},
and useful information on quark and gluon angular momentum contributions to the proton spin \cite{Yu:2024ovn}.

Here, we reiterate the key tenets of the AO scheme.
(\emph{a}) There is an effective charge, $\alpha_{1\ell}(k^2)$, of the type explained in Refs.\,\cite{Grunberg:1980ja, Grunberg:1982fw} and reviewed in Ref.\,\cite{Deur:2023dzc}, that, when used to integrate the leading-order perturbative DGLAP equations, defines an evolution scheme for all parton DFs that is all-orders exact.
The form of $\alpha_{1\ell}(k^2)$ is largely immaterial.  Nevertheless, the process-independent (PI) charge described in Refs.\,\cite{Binosi:2014aea, Binosi:2016nme, Cui:2019dwv} has all required properties.
(\emph{b}) At the hadron scale, $\zeta_{\cal H}<m_p$, all properties of a given hadron are carried by its valence degrees of freedom.  So, at this scale, DFs associated with glue and sea quarks are zero.  Nonzero values for glue and sea DFs are obtained via AO evolution to $\zeta > \zeta_{\cal H}$.

In principle, it is not necessary to specify the value of $\zeta_{\cal H}$ when employing AO evolution.  Nevertheless, if a particular effective charge is chosen, then the value becomes known.
The PI charge calculated in Ref.\,\cite{Cui:2019dwv} defines a screening mass, whose value is a natural choice for the hadron-scale:
\begin{equation}
\zeta_{\cal H}=0.331(2)\,{\rm GeV}.
\label{zetaH}
\end{equation}
Analysis of results from lQCD relating to the pion valence quark DF yields a consistent value \cite{Lu:2023yna}: $\zeta_{\cal H} = 0.350(44)\,{\rm GeV}$.

\subsection{Momentum conservation and glue}
\label{MomGlue}
In DF evolution, parton momentum conservation is automatic.
For FFs, however, the off-diagonal terms in the matrix of splitting functions are interchanged, in consequence of which the singlet FFs pass momentum into the gluon FFs, with the loss and gain being unbalanced -- see, \emph{e.g}., Ref.\,\cite[Eqs.\,(21), (22)]{Xing:2023pms}.
Notwithstanding this, FF evolution ensures that flavour is conserved during hadronisation:
\begin{equation}
\int_0^1 dz \, D_{N_{\mathpzc q}}^h(z;\zeta) \stackrel{\zeta > \zeta_{\cal H}}{=}
\int_0^1 dz \, D_{N_{\mathpzc q}}^h(z;\zeta_{\cal H})\,.
\end{equation}

When inferring FFs through fits to data, momentum conservation can be enforced by requiring that the input FFs for each parton produce a collection of first Mellin moments whose sum is unity after all final-state hadrons are included -- see, \emph{e.g}., Ref.\,\cite[Eq.\,(11)]{Hirai:2007cx}.  
%
%
However, there are issues with implementing that constraint in such fits.
(\emph{i}) No experiment can reach $z=0$.  
(\emph{ii})  For a given set of data, the empirical hadron final-state space is unknown because hadrons in addition to $\pi$, $K$, $p$, $\bar p$ might be produced and yet go undetected.
Thus, contemporary FF fit procedures are typically only able to ensure that the results are consistent with some set of practitioner-dependent momentum sum rule bounds or inequalities.  
Such subjective ambiguities are avoided in our approach because: the predictions are delivered on the entire $z$ domain; and we limit the hadron final-state space, making comparisons between predictions and data that are constrained to lie within the same state space.  

In adapting the AO scheme, Ref.\,\cite{Xing:2023pms} observed that if, for instance, one begins with $D_g^h(x;\zeta_{\cal H})\equiv 0$, then evolution takes momentum from $D_S^h(x;\zeta_{\cal H})$, feeding it into $D_g^h(x;\zeta_{\cal H})$.  Overall, however, momentum is lost to the unresolved parton shower.

If one instead assumes $D_g^h(z;\zeta_{\cal H})\neq 0$, then there is always a value of
\begin{equation}
\langle z\rangle_{D_g^h}^{\zeta_{\cal H}} =
\sum_{q=u,d,s} \int_0^1 dz z D_{g_{\mathpzc q}}^h(z;\zeta_{\cal H})\,,
\end{equation}
where $D_{{g}_{\mathpzc q}}^h(z;\zeta_{\cal H})$ is the hadron-scale gluon FF that mixes with the $q$ valence quark, such that
\begin{align}
\sum_{q=u,d,s,\ldots} & \langle z \rangle_{D_{S_{\mathpzc q}}^h}^{\zeta}
 + \langle z \rangle_{D_{{g}}^h}^{\zeta} \nonumber \\
%
& \stackrel{\forall \zeta>\zeta_{\cal H}}{=}
\sum_{q=u,d,s}
\langle z \rangle_{D_{S_{\mathpzc q}}^h}^{\zeta_{\cal H}}
+ \langle z\rangle_{D_g^h}^{\zeta_{\cal H}},
%
%
\label{MomBalance}
\end{align}
where the sum in the first line ranges over all quarks that can be produced at the given $\zeta>\zeta_{\cal H}$.
%
Considering FF evolution equations with splitting functions defined for $n_f$ massless (evolution-active) quarks, the critical value of the momentum fraction distributed by the gluon FF is \cite{Xing:2023pms}:
\begin{equation}
\label{GlueMomFraction}
\langle z\rangle_{D_g^h}^{\zeta_{\cal H}} = 1/[1+2 n_f]\,.
\end{equation}

This discussion means that the singlet form of \linebreak Eq.\,\eqref{JetEq} is incomplete.  Each singlet jet equation in QCD should involve gluon contributions to the cascade, because of $g \leftrightarrow q + \bar q$ mixing, and also, therefore, heavier quark + antiquark pairs, albeit to a lesser extent.   Following Ref.\,\cite{Xing:2023pms}, we implement this phenomenologically by writing\\[-3ex]
{\allowdisplaybreaks
\begin{subequations}
\label{GlueAnsatz}
\begin{align}
D_{S_{\mathpzc q}}^h(z;\zeta_{\cal H}) &
\to \tilde D_{S_{\mathpzc q}}^h(z;\zeta_{\cal H})
+ \tilde D_{g_{\mathpzc q}}^h(z;\zeta_{\cal H}) \\
& = (1-\delta) D_{S_{\mathpzc q}}^h(z;\zeta_{\cal H})
+ \delta D_{g_{\mathpzc q}}^h(z;\zeta_{\cal H})\,,
\end{align}
\end{subequations}
with the constant $\delta\in (0,1 )$ chosen to guarantee Eq.\,\eqref{MomBalance}
and
\begin{align}
D_{g_{\mathpzc q}}^h(z;\zeta_{\cal H}) &
\propto [D_{S_{\mathpzc q}}^h(z;\zeta_{\cal H})-D_{N_{\mathpzc q}}^h(z;\zeta_{\cal H})]\,,
\label{DgDSpi}
\end{align}
normalised to ensure
\begin{equation}
\int_0^1dz\,z\, D_{g_{\mathpzc q}}^h(z;\zeta_{\cal H})
= \int_0^1dz\,z\,D_{S_{\mathpzc q}}^h(z;\zeta_{\cal H}) \,.
\end{equation}
With Eq.\,\eqref{DgDSpi}, one has a minimal \emph{Ansatz}: the glue FF profile for each quark flavour matches the pointwise behaviour of the unfavoured $q\to h$ FF.  As will be seen below, this is sufficient to achieve the desired outcome.
}

In order to be explicit concerning momentum conservation, it is furthermore convenient to expand the FF evolution equations as follows.
As usual, write
$\breve D(z) = z D(z)$; then
\begin{equation}
\breve D_{S_{\mathpzc q}}^h(z;\zeta) = \sum_{{\mathpzc q}^\prime=u,d,s,c} \breve
D_{S_{\mathpzc q}^{{\mathpzc q}^\prime}}^h(z;\zeta)\,,
\end{equation}
where $D_{S_{\mathpzc q}^{{\mathpzc q}^\prime}}^h(z;\zeta)$ describes the evolution-induced chain ${\mathpzc q} \to_{\rm evolution} {\mathpzc q}^\prime \to_{\rm fragmentation} h$, \emph{i.e}., ${\mathpzc q}^\prime$ fragments into $h$ by delivering the momentum it took from ${\mathpzc q}$;
and subsequently solve the associated tower of coupled evolution equations [$t=\ln \zeta^2$]
\begin{subequations}
\label{DGLAPsinglet}
\begin{align}
    \frac{d}{dt} \breve D_{S_{\mathpzc q}^{{\mathpzc q}^\prime}}^h(z;t) &
    = \frac{\alpha(t)}{2\pi} \int_z^1\,dy
    \left[ P_{q^\prime q}(y) \breve D_{S_{\mathpzc q}^{{\mathpzc q}^\prime}}^h(z/y;t) \right. \nonumber \\
    & \qquad \left. + 2 P_{gq^\prime}(y) \breve D_{g_{\mathpzc q}}^h(z/y;t) \right] ,\\
    \frac{d}{dt} \breve D_{g_{\mathpzc q}}^h(z;t) & = \frac{\alpha(t)}{2\pi} \int_z^1\,dy
    \left[ \sum_{{\mathpzc q}^\prime} P_{q^\prime g}(y)
    \breve D_{S_{\mathpzc q}^{{\mathpzc q}^\prime}}^h(z/y;t) \right. \nonumber \\
    & \qquad \left. + P_{gg} \breve D_{g_{\mathpzc q}}^h(z/y;t) \right] ,
\end{align}
\end{subequations}
where the massless splitting functions, which don't distinguish between $q$, $q^\prime$, are given in Ref.\,\cite[Eq.\,(14)]{Xing:2023pms}.  These equations should be solved with the initial conditions
\begin{subequations}
\label{EvolutionInitial}
\begin{align}
D_{S_{\mathpzc u}^{{\mathpzc u}}}^\pi(z;\zeta_{\cal H}) & = D_{S_{\mathpzc u}}^\pi(z;\zeta_{\cal H})\,,
\;
D_{S_{\mathpzc u}^{{\mathpzc q}^\prime \neq u}}^\pi(z;\zeta_{\cal H}) = 0\,,\\
D_{S_{\mathpzc u}^{{\mathpzc u}}}^{K^+}(z;\zeta_{\cal H}) & = D_{S_{\mathpzc u}}^{K^+}(z;\zeta_{\cal H})\,,
\;
D_{S_{\mathpzc u}^{{\mathpzc q}^\prime \neq u}}^{K^+}(z;\zeta_{\cal H}) = 0\,,
\end{align}
\end{subequations}
etc.

%

It is worth observing that $D_g^h(z;\zeta_{\cal H})\neq 0$ does not mark a deviation from the standard AO evolution principle that $\zeta_{\cal H}$ is the scale at which all properties of a given hadron are carried by its valence (quasiparticle) degrees-of-freedom \cite{Yin:2023dbw}.
This is plain once one notes that, following a given collision, the fragmentation process inserts one of the produced quasiparticle partons into a particular final-state hadron; but, irrespective of the scale, not all the collision debris can correspond to a valence degree of freedom in that hadron.
%

\begin{figure*}[t]
\hspace*{-1ex}
\begin{tabular}{ccc}
{\sf A} & {\sf B} & {\sf C} \\
\includegraphics[clip, width=0.655\columnwidth]{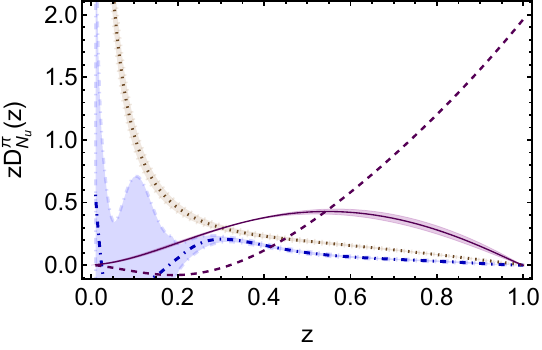} &
\includegraphics[clip, width=0.655\columnwidth]{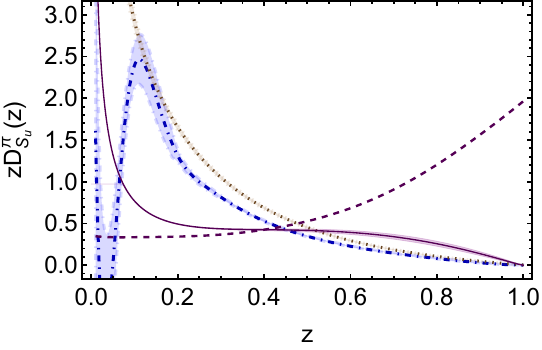} &
\includegraphics[clip, width=0.655\columnwidth]{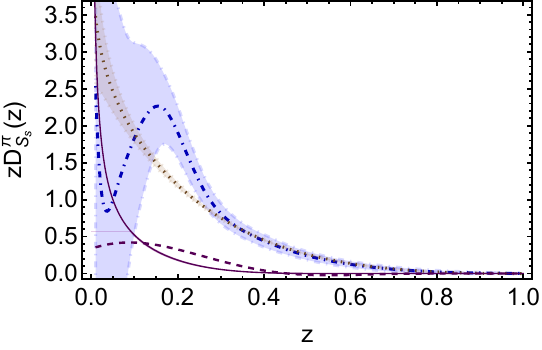}
\end{tabular}
\begin{tabular}{ccc}
{\sf D} & {\sf E} &  \\
\includegraphics[clip, width=0.655\columnwidth]{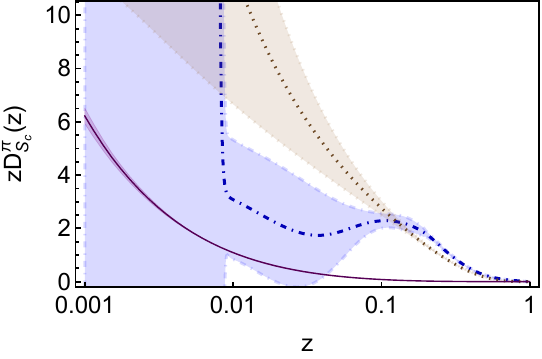} &
\includegraphics[clip, width=0.655\columnwidth]{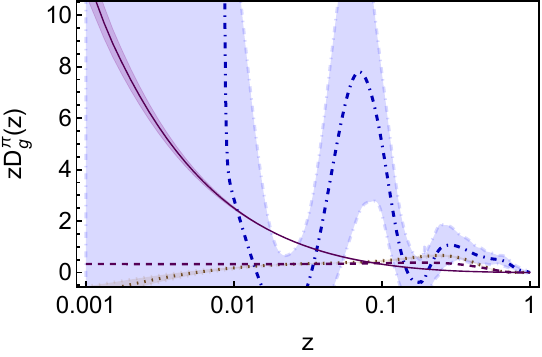} &
\\
\end{tabular}
\caption{\label{FSCIJetpi}
SCI results for pion fragmentation functions, defined in Eqs.\,\eqref{Spi}, \eqref{Npi}.
Solutions of cascade equations, Eq.\,\eqref{JetExplicit} -- dashed purple curves.
AO evolution of those curves to $\zeta=\zeta_2 := 2\,$GeV -- solid purple curves, with uncertainty bands obtained as described in Sect.\,\ref{HSuncertainty}.
Comparison curves are inferences from:
high-energy lepton-lepton, lepton-hadron and hadron-hadron scattering data \cite[JAM]{Moffat:2021dji} -- dotted brown curves, within like coloured bands;
and electron-positron annihilation and lepton-nucleon semi-inclusive deep-inelastic scattering data \cite[MAPFF]{AbdulKhalek:2022laj} -- dot-dashed blue curves within like-coloured bands.
}
\end{figure*}

\begin{figure*}[t]
\hspace*{-1ex}\begin{tabular}{ccc}
{\sf A} & {\sf B} & {\sf C} \\
\includegraphics[clip, width=0.655\columnwidth]{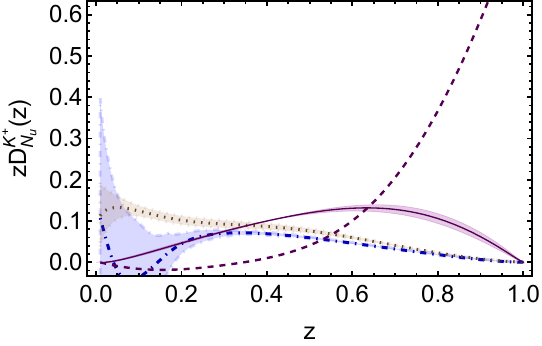} &
\includegraphics[clip, width=0.655\columnwidth]{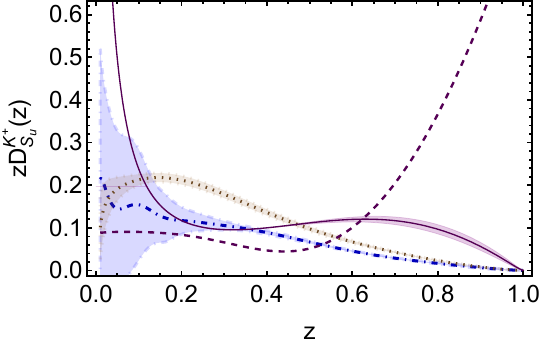} &
\includegraphics[clip, width=0.655\columnwidth]{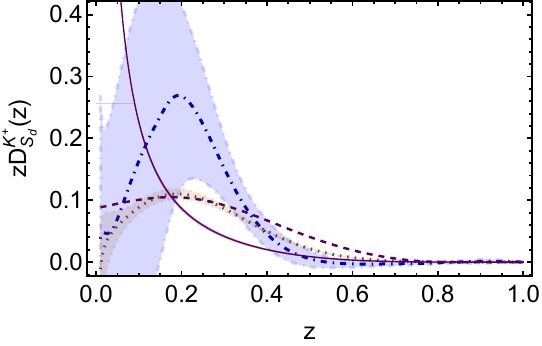}
\end{tabular}
\begin{tabular}{ccc}
{\sf D} & {\sf E} & \\
\includegraphics[clip, width=0.655\columnwidth]{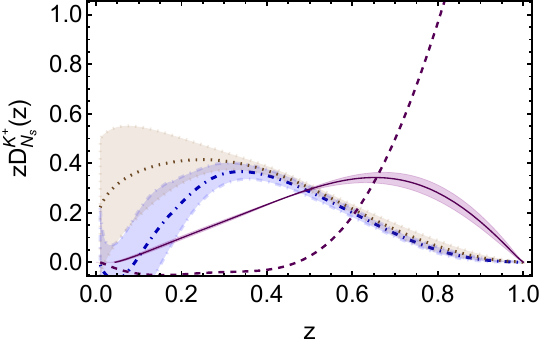} &
\includegraphics[clip, width=0.655\columnwidth]{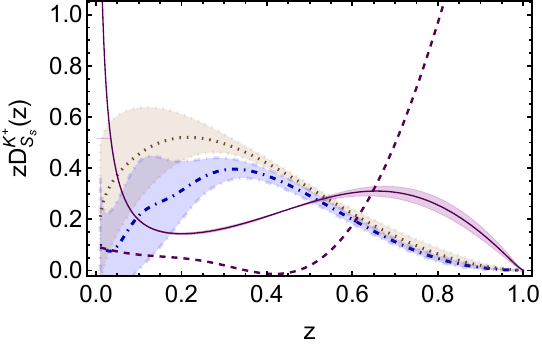} & \\
\end{tabular}
\\
\begin{tabular}{ccc}
{\sf F} & {\sf G} & \\
\includegraphics[clip, width=0.655\columnwidth]{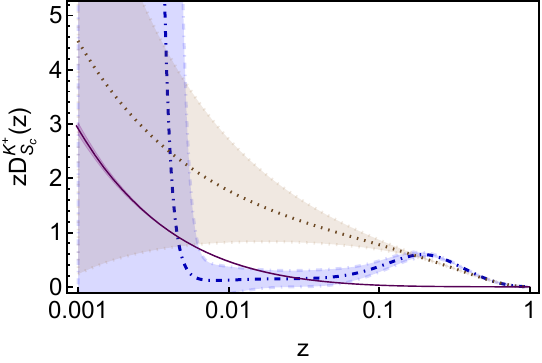} &
\includegraphics[clip, width=0.655\columnwidth]{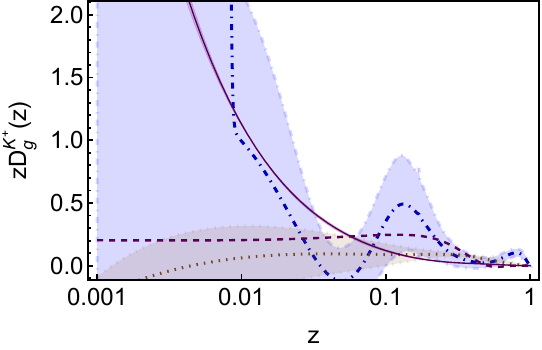} & \\
\end{tabular}
\caption{\label{FSCIJetK}
SCI results for kaon fragmentation functions, defined in Eqs.\,\eqref{SK}\,--\,\eqref{NKs}.
Solutions of cascade equations, Eq.\,\eqref{JetExplicit} -- dashed purple curves.
AO evolution of those curves to $\zeta=\zeta_2 := 2\,$GeV -- solid purple curves, with uncertainty bands obtained as described in Sect.\,\ref{HSuncertainty}.
Comparison curves are inferences from:
high-energy lepton-lepton, lepton-hadron and hadron-hadron scattering data \cite[JAM]{Moffat:2021dji} -- dotted brown curves, within like coloured bands;
and electron-positron annihilation and lepton-nucleon semi-inclusive deep-inelastic scattering data \cite[MAPFF]{AbdulKhalek:2022laj} -- dot-dashed blue curves within like-coloured bands.
}
\end{figure*}

\section{Systematic Uncertainties}
\label{SysUnc}
\subsection{EFFs}
\label{sectEFFs}
As stated at the outset, herein we deliver FF predictions obtained using two different sets of EFFs.  The first is based on the SCI and the second follows from realistic CSM predictions for pion and kaon DFs.

Regarding the SCI EFFs, we note that the SCI is not a precision tool, but it does have many merits, \emph{e.g}.:
algebraic simplicity;
simultaneous applicability to a large array of systems and processes;
potential for delivering insights that link and explain numerous phenomena;
and capability to serve as a tool for validating algorithms employed in computations that depend upon high performance computing.
Notably, today's SCI applications are typically parameter-free.
Herein as elsewhere, the SCI framework delivers an illustrative ``rule of thumb'' guide as to what one should expect for a given set of physical quantities.
Recent studies, pertinent to our discussion, are described in Refs.\,\cite{Xing:2023pms, Yu:2024ovn, Yu:2025fer, Cheng:2024gyv}.

On the other hand, the CSM EFFs are realistic because, wherever sensible comparisons are possible, the source CSM DFs agree with available empirical inferences and/or results obtained in contemporary lQCD simulations; see, \emph{e.g}., Refs.\,\cite{Lu:2022cjx, Lu:2023yna, Xu:2024nzp, Alexandrou:2024zvn}.
With the CSM DFs thus benchmarked against experiment and theory, no uncertainty is attached to the choice of EFFs herein: those obtained using CSM inputs provide the physical foundation.

It is worth stressing that one should not, therefore, view subsequent comparisons between SCI and CSM results as providing any sort of uncertainty estimate (except in one case, discussed further below).  Instead, the fact that we will find SCI and CSM results to be in semiquantitative agreement is a further validation of SCI utility, not any check on CSM veracity.

\subsection{Hadron scale}
\label{HSuncertainty}
In many existing studies using AO evolution, DF and FF predictions are reported along with an uncertainty obtained by varying the value of the hadron scale by $\pm 5$\%; see, \emph{e.g}., Refs.\,\cite{Lu:2022cjx, Xing:2023pms, Yu:2024ovn}.
The results therein show that the impact of such variation is small on individual evolved DFs and FFs and cancels in all ratios.
Notwithstanding that, we follow these examples, \emph{viz}.\ all subsequent results for evolved FFs are reported along with an uncertainty band obtained by beginning with $\zeta_{\cal H}$ in Eq.\,\eqref{zetaH} and varying this value according to $\zeta_{\cal H} \to \zeta_{\cal H} (1\pm 0.05)$.

It is natural to ask whether $\pm 5$\% is reasonable.  An answer is already apparent in Eq.\,\eqref{zetaH}.  Namely, this value of the hadron scale is a prediction derived from the process-independent QCD effective charge calculated in Ref.\,\cite{Cui:2019dwv} and its associated uncertainty is merely $0.6$\%.  So, a $\pm 5$\% variation provides a very conservative expression of the underlying uncertainty.

Furthermore, as also mentioned after Eq.\,\eqref{zetaH}, inferences of $\zeta_{\cal H}$ from analyses of a range of lQCD DF studies \cite{Lu:2023yna}, which is a different approach from that used Ref.\,\cite{Cui:2019dwv}, yields a consistent value, with an uncertainty of 13\%.  So, if desired, the choice $\pm ±5$\% may be viewed as (practically) the mean of the uncertainties associated with two distinct methods of determining $\zeta_{\cal H}$.
Altogether, therefore, the hadron-scale uncertainty that we provide herein is conservative; hence, reasonable.

\subsection{Gluon profile}
\label{glueFFSuncertainty}
The remaining potential source of systematic uncertainty in our predictions is the \emph{Ansatz} for the hadron-scale gluon FF, which is explained in Sect.\,\ref{MomGlue} and introduced explicitly in Eq.\,\eqref{DgDSpi}.  In evaluating the possible impact of this choice, one should consider just what functional forms are physically reasonable.

The first point is that, irrespective of its $z$-profile, the first moment of $D_{g_{\mathpzc q}}^h(z;\zeta_{\cal H})$ is fixed; see Eq.\,\eqref{GlueMomFraction}.  This is a valuable constraint on the normalisation.

Second, since glue cannot produce a valence quark to begin the hadronisation process, $D_{g_{\mathpzc q}}^h(z;\zeta_{\cal H})$ must have the profile of an unfavoured FF.

So, one seeks a $z$-moment constrained \emph{Ansatz} for $D_{g_{\mathpzc q}}^h(z;\zeta_{\cal H})$ that has the profile of an unfavoured FF and which is internally consistent with the calculational framework.
This last observation means that one cannot sensibly choose an arbitrary function whose form is unrelated to the underlying interaction.  The only simple function that fulfils these requirements is Eq.\,\eqref{DgDSpi}.
We thus arrive at the sole case where a comparison between SCI and CSM predictions does provide a sensible means of judging a systematic uncertainty, \emph{viz}.\ that introduced by the Eq.\,\eqref{DgDSpi} \emph{Ansatz}.  We exploit this below.


\section{SCI Fragmentation Functions}
\label{SecSCIFFs}
Solving the jet cascade equations using SCI EFF inputs to complete Eqs.\,\eqref{JetExplicit}, one obtains the dashed purple curves in Fig.\,\ref{FSCIJetpi}\,A-C and  Fig.\,\ref{FSCIJetK}\,A-E.
Subsequently evolving those results, according to the procedure explained in Sect.\,\ref{FFEvolution},
including $s$ and $c$ quark mass thresholds -- see, \emph{e.g}., Ref.\,\cite[Sec.\,2]{Lu:2022cjx},
and setting
$\delta=0.11$;
then one obtains the SCI predictions for $D_{S_{\mathpzc q},N_{\mathpzc q}}^h(z;\zeta_2=2\,{\rm GeV})$ drawn in Figs.\,\ref{FSCIJetpi}, \ref{FSCIJetK} (solid purple curves).
%
For comparison, we have drawn the inferences from data reported in Refs.\,\cite{Moffat:2021dji, AbdulKhalek:2022laj}.  Plainly, they are mutually incompatible on $z\lesssim 0.5$.
To assist with image clarity, we do not include the phenomenological fits from Ref.\,\cite{Gao:2024dbv}, but they also diverge widely from the fits in Refs.\,\cite{Moffat:2021dji, AbdulKhalek:2022laj}
Simply put, phenomenology available today does not deliver objective FF results: the results obtained are practitioner dependent.
\begin{table}[t]
\caption{\label{MomFracs}
SCI FF momentum fractions obtained from solutions of the cascade equations at the hadron scale and after evolution to $\zeta = \zeta_2 := 2\,$GeV, following the prescription described in Sect.\,\ref{FFEvolution}.
(No entry means the fraction is zero.  $c\to q \to h$ contributions are negligible in all cases.)
 }
\begin{center}
\begin{tabular*}
{\hsize}
{
l@{\extracolsep{0ptplus1fil}}
|l@{\extracolsep{0ptplus1fil}}
l@{\extracolsep{0ptplus1fil}}
|l@{\extracolsep{0ptplus1fil}}
l@{\extracolsep{0ptplus1fil}}
l@{\extracolsep{0ptplus1fil}}|}\hline
$h\ $ & \multicolumn{2}{c|}{$\pi^+ + \pi^0+\pi^-$} & \multicolumn{2}{c}{$K^+$} \\ \hline
 & $\zeta_{\cal H}\ $ & $\zeta_{2}\ $ & $\zeta_{\cal H}\ $ & $\zeta_2\ $ \\
$\langle z\rangle_{D_{S_{\mathpzc u}^{\mathpzc u}}}^h\ $ & $0.664\ $  & $0.433\ $  & $0.182\ $  & $0.119\ $ \\
$\langle z\rangle_{D_{S_{\mathpzc u}^{\mathpzc d}}}^h\ $ &  & $0.115\ $  &  & $0.032\ $ \\
$\langle z\rangle_{D_{S_{\mathpzc u}^{\mathpzc s}}}^h\ $ &  & $0.085\ $  &  & $0.023\ $ \\
$\langle z\rangle_{D_{S_{\mathpzc u}^{\mathpzc c}}}^h\ $ &  & $0.031\ $  &  & $0.009\ $ \\
$\langle z\rangle_{D_{S_{\mathpzc d}^{\mathpzc u}}}^h\ $ &   & $0.115\ $  &  & $0.007\ $ \\
$\langle z\rangle_{D_{S_{\mathpzc d}^{\mathpzc d}}}^h\ $ & $0.664\ $  & $0.443\ $  & $0.042\ $ & $0.028\ $ \\
$\langle z\rangle_{D_{S_{\mathpzc d}^{\mathpzc s}}}^h\ $ & & $0.085\ $  & & $0.005\ $ \\
$\langle z\rangle_{D_{S_{\mathpzc d}^{\mathpzc c}}}^h\ $ & & $0.031\ $  & & $0.002\ $ \\
$\langle z\rangle_{D_{S_{\mathpzc s}^{\mathpzc u}}}^h\ $ & & $0.017\ $  & & $0.069\ $ \\
$\langle z\rangle_{D_{S_{\mathpzc s}^{\mathpzc d}}}^h\ $ & & $0.017\ $  & & $0.069\ $ \\
$\langle z\rangle_{D_{S_{\mathpzc s}^{\mathpzc s}}}^h\ $ & $0.098\ $  & $0.059\ $  & $0.396\ $ & $0.239\ $ \\
$\langle z\rangle_{D_{S_{\mathpzc s}^{\mathpzc c}}}^h\ $ & & $0.005\ $  & & $0.019\ $ \\
%
$\langle z\rangle_{D_{{g}_{\mathpzc u}}^h}^\zeta\ $ & $0.083\ $  & 0.083 & $0.023\ $ & 0.023 \\
$\langle z\rangle_{D_{{g}_{\mathpzc d}}^h}^\zeta\ $ & $0.083\ $  & 0.083 & $0.005\ $ & 0.005 \\
$\langle z\rangle_{D_{{g}_{\mathpzc s}}^h}^\zeta\ $ & $0.012\ $  & $0.012$ & $0.050\ $  & $0.050\ $ \\
%
%
%
\hline
\end{tabular*}
\end{center}
\end{table}

{\allowdisplaybreaks
We will postpone a discussion of the compatibility of the phenomenological data fits with our predictions until we describe realistic CSM results below.  Here we focus on momentum conservation, illustrating the impacts of $D_g^h(z;\zeta_{\cal H})\neq 0$.
At $\zeta_{\cal H}$, the cascade solution FFs in Figs.\,\ref{FSCIJetpi}, \ref{FSCIJetK} yield the momentum fractions in Table~\ref{MomFracs}-columns~1, 3 -- see page~\pageref{MomFracs}, where
\begin{equation}
\langle z\rangle_{D^h}^\zeta
= \int_0^1 dz\,z\, D^h(z;\zeta)\,.
\end{equation}
Note now that
\begin{align}
& \sum_{\rm all~h}\int_0^1dz\,z\,\left[
D_{S_{\mathpzc u}}^h(z;\zeta_{\cal H}) +
D_{{g}_{\mathpzc u}}^h(z;\zeta_{\cal H})\right] \nonumber \\
& =\;
\stackrel{u\to \pi}{0.664}
+ \stackrel{u\to K^+ K^-}{0.182}
+ \stackrel{u\to K^0 \bar K^0}{0.042}  \nonumber \\
& \qquad + \stackrel{u(g)\to \pi}{0.083} + \stackrel{u(g)\to K^+ K^-}{0.023} +
\stackrel{u(g)\to K^0 \bar K^0}{0.005}  = 1.0\,.
\label{MomConsAllhu}
\end{align}
Here we have used Eqs.\,\eqref{GParity} to identify, \emph{e.g}., $d\to K^+ K^-$ with $u\to K^0 \bar K^0$.

After evolution, the momentum is partitioned more widely, with the results listed in Table~\ref{MomFracs}-columns~2, 4 -- see page~\pageref{MomFracs}.  In this case:
\begin{align}
& \sum_{\rm all~h} \int_0^1dz\,z\,\left[
\sum_{\mathpzc q}D_{S_{\mathpzc u}^{\mathpzc q}}^h(z;\zeta_{2}) +
D_{{g}_{\mathpzc u}}^h(z;\zeta_{2})\right] \nonumber \\
& =\;
\stackrel{u\to u\to  \pi}{0.433}
\quad + \stackrel{u\to d\to  \pi}{0.115}
+ \stackrel{u\to s\to  \pi}{0.085}
+ \stackrel{u\to c\to  \pi}{0.031} \nonumber \\
&
\quad + \stackrel{u\to u\to  K^{+ -}}{0.119}
+ \stackrel{u\to d \to K^{+ -}}{0.032}
+ \stackrel{u\to s \to K^{+ -} }{0.023}
+ \stackrel{u\to c \to K^{+ -}}{0.009} \nonumber \\
&
\quad + \stackrel{d\to u \to K^{+ -}}{0.007}
+ \stackrel{d\to d \to K^{+ -}}{0.028}
+ \stackrel{d\to s \to K^{+ -}}{0.005}
+ \stackrel{d\to c \to K^{+ -}}{0.002}
\nonumber \\
&
\quad + \stackrel{u(g)\to \pi}{0.083} + \stackrel{u(g)\to K^+ K^-}{0.023} +
\stackrel{u(g)\to K^0 \bar K^0}{0.005}  \label{Line3} \\
& = 1.0\,.
\end{align}
In the third line on the right-hand side of Eq.\,\eqref{Line3}, we again used $ u\to K^0 \bar K^0 \leftrightarrow d\to K^+ K^- $.

Using Table~\ref{MomFracs}, one may also check momentum conservation for the $s$ quark.  First, at the hadron scale:
\begin{align}
& \sum_{\rm all~h}\int_0^1dz\,z\,\left[
D_{S_{\mathpzc s}}^h(z;\zeta_{\cal H}) +
D_{{g}_{\mathpzc s}}^h(z;\zeta_{\cal H})\right] \nonumber \\
& =\;
\stackrel{s\to \pi}{0.098}
+ \stackrel{s\to K^+ K^-}{0.396}
+ \stackrel{s\to K^0 \bar K^0}{0.396}  \nonumber \\
& \qquad + \stackrel{s(g)\to \pi}{0.012} + \stackrel{s(g)\to K^+ K^-}{0.050} +
\stackrel{s(g)\to K^0 \bar K^0}{0.050}  = 1.0\,,
\end{align}
where we have used $s\to K^+ K^- \leftrightarrow s\to K^0 \bar K^0$ -- see Eq.\,\eqref{sKpm00}.
Then, after evolution:
\begin{align}
& \sum_{\rm all~h} \int_0^1dz\,z\,\left[
\sum_{\mathpzc q}D_{S_{\mathpzc s}^{\mathpzc q}}^h(z;\zeta_{2}) +
D_{{g}_{\mathpzc s}}^h(z;\zeta_{2})\right] \nonumber \\
& =\;
\stackrel{s\to u\to  \pi}{0.017}
\quad
+ \stackrel{s\to d\to  \pi}{0.017}
+ \stackrel{s\to s\to  \pi}{0.059}
+ \stackrel{s\to c\to  \pi}{0.005} \nonumber \\
&
\;\;\, + 2 [
   \stackrel{s\to u\to  K^{+ -}}{0.069}
+ \stackrel{s\to d \to K^{+ -}}{0.069}
+ \stackrel{s\to s \to K^{+ -} }{0.239}
+ \stackrel{s\to c \to K^{+ -}}{0.019} ] \nonumber \\
%
&
\quad + \stackrel{s(g)\to \pi}{0.012} + \stackrel{s(g)\to K^+ K^-}{0.050} +
\stackrel{s(g)\to K^0 \bar K^0}{0.050}   \\
& = 1.0\,.
\label{MomConsAllhs}
\end{align}
}

Evidently, in all cases, the individual gluon momentum fractions are preserved under evolution; hence, as found in Ref.\,\cite{Xing:2023pms}, so is the total.

\begin{figure}[t]
\leftline{\large\sf A}
\vspace*{-2ex}

\centerline{%
\includegraphics[clip, width=0.95\linewidth]{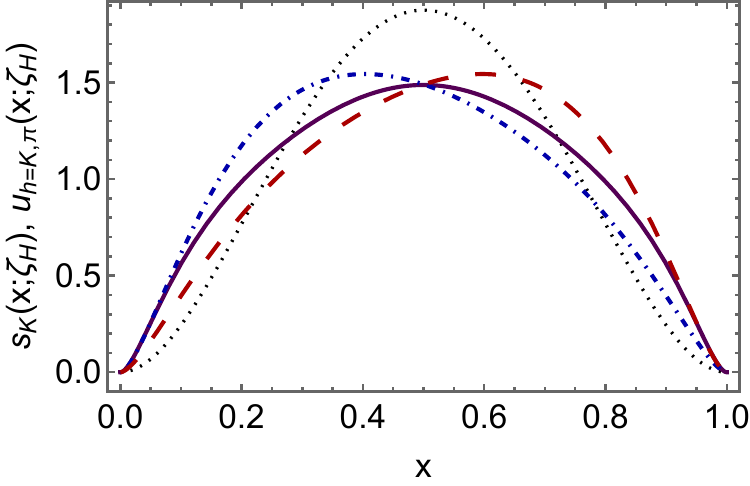}}

\vspace*{1ex}

\leftline{\large\sf B}
\vspace*{-2ex}

\centerline{%
\hspace*{-1ex}\includegraphics[clip, width=0.96\linewidth]{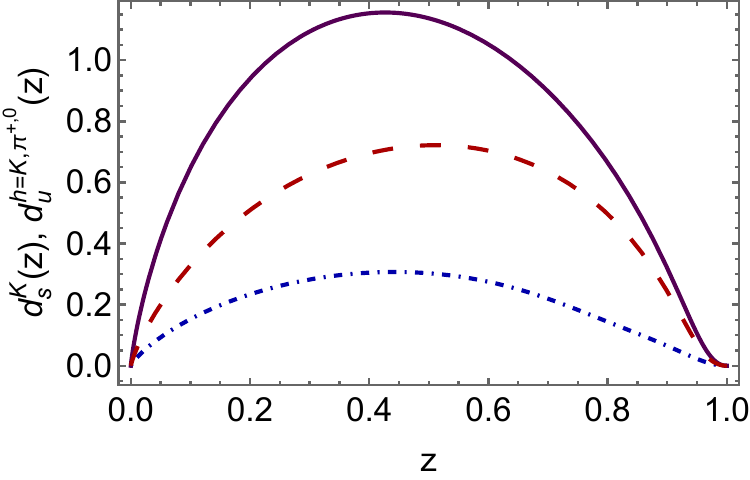}}
%

%
\caption{\label{CSMpiKDFs}
{\sf Panel A}.
Dressed valence quark parton distribution functions evaluated using CSMs in Ref.\,\cite{Cui:2020tdf}:
${\mathpzc s}_{K^-}(x ; \zeta_H)$ -- long-dashed red curve;
${\mathpzc u}_{K^+}(x ; \zeta_H)$ -- dot-dashed blue;
${\mathpzc u}_{\pi^+}(x ; \zeta_H)$ -- solid purple;
scale-free DF in Eq.\,\eqref{ScaleFree} -- dotted black.
{\sf Panel B}.
Realistic elementary fragmentation functions, obtained from the $\pi, K$ curves in {\sf Panel A} using Eqs.\,\eqref{DLYR}.
$d_{\mathpzc s}^{K^-}(x ; \zeta_H)$ -- long-dashed red curve;
$d_{\mathpzc u}^{K^+}(x ; \zeta_H)$ -- dot-dashed blue ;
$d_{\mathpzc u}^{\pi^++\pi^0}(x ; \zeta_H)$ -- solid purple.
}
\end{figure}

\begin{figure*}[t]
\hspace*{-1ex}
\begin{tabular}{ccc}
{\sf A} & {\sf B} & {\sf C} \\
\includegraphics[clip, width=0.655\columnwidth]{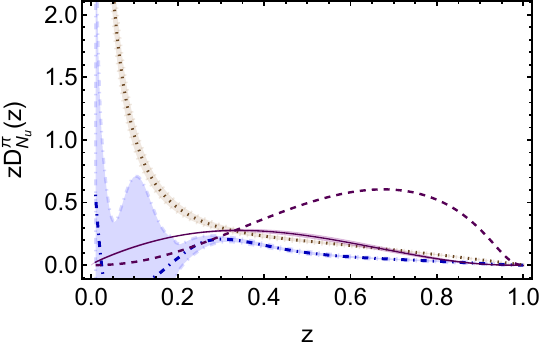} &
\includegraphics[clip, width=0.655\columnwidth]{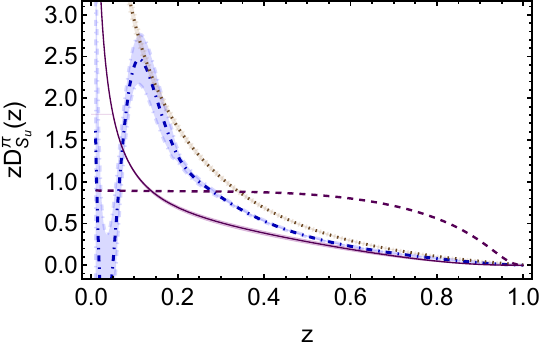} &
\includegraphics[clip, width=0.655\columnwidth]{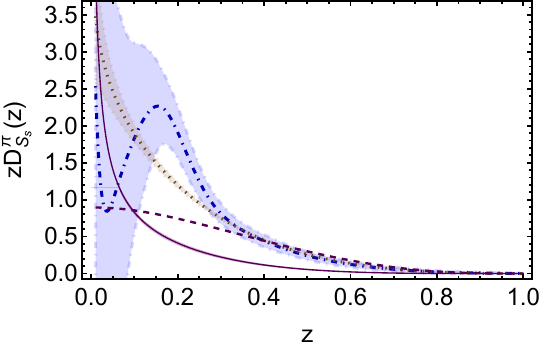}
\end{tabular}
\begin{tabular}{ccc}
{\sf D} & {\sf E} &  \\
\includegraphics[clip, width=0.655\columnwidth]{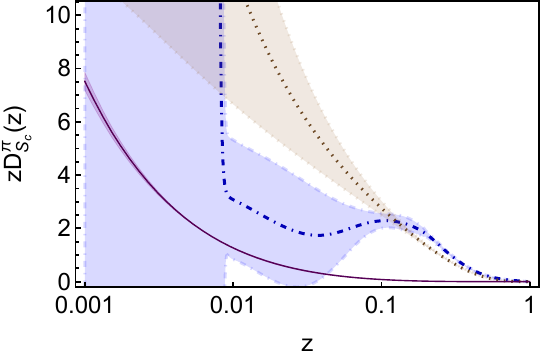} &
\includegraphics[clip, width=0.655\columnwidth]{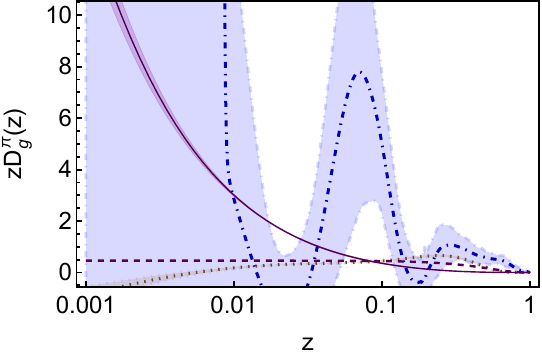} &
\\
\end{tabular}
\caption{\label{FCSMJetpi}
CSM results for pion fragmentation functions, defined in Eqs.\,\eqref{Spi}, \eqref{Npi}.
Solutions of cascade equations, Eq.\,\eqref{JetExplicit} -- dashed purple curves.
AO evolution of those curves to $\zeta=\zeta_2 := 2\,$GeV -- solid purple curves, with uncertainty bands obtained as described in Sect.\,\ref{HSuncertainty}.
Comparison curves are inferences from:
high-energy lepton-lepton, lepton-hadron and hadron-hadron scattering data \cite[JAM]{Moffat:2021dji} -- dotted brown curves, within like coloured bands;
and electron-positron annihilation and lepton-nucleon semi-inclusive deep-inelastic scattering data \cite[MAPFF]{AbdulKhalek:2022laj} -- dot-dashed blue curves within like-coloured bands.
}
\end{figure*}

\begin{figure*}[t]
\hspace*{-1ex}
\begin{tabular}{ccc}
{\sf A} & {\sf B} & {\sf C} \\
\includegraphics[clip, width=0.655\columnwidth]{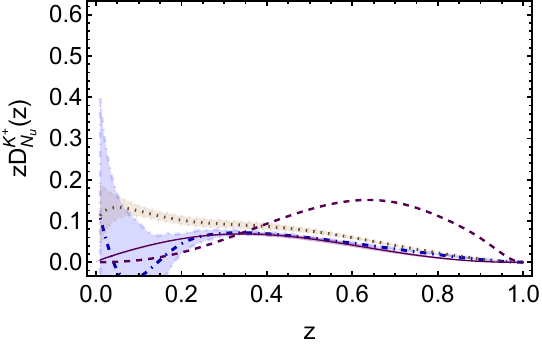} &
\includegraphics[clip, width=0.655\columnwidth]{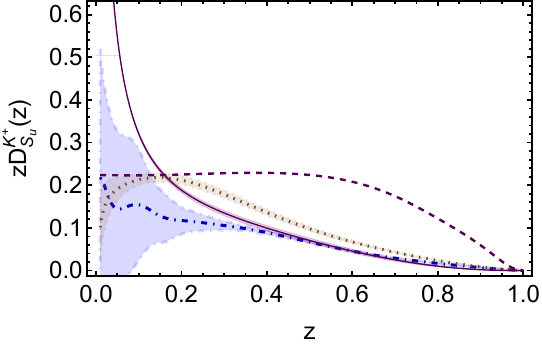} &
\includegraphics[clip, width=0.655\columnwidth]{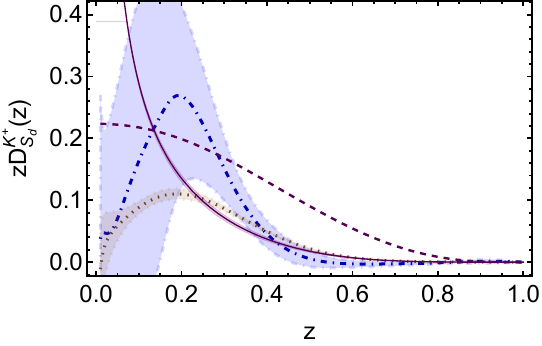}
\end{tabular}
\begin{tabular}{ccc}
{\sf D} & {\sf E} & \\
\includegraphics[clip, width=0.655\columnwidth]{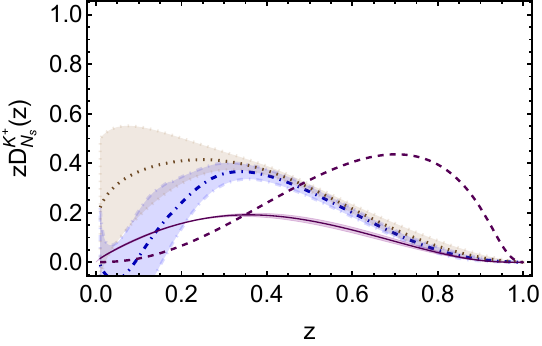} &
\includegraphics[clip, width=0.655\columnwidth]{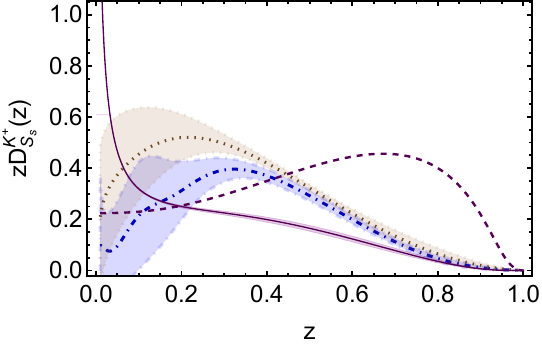} & \\
\end{tabular}
\\
\begin{tabular}{ccc}
{\sf F} & {\sf G} & \\
\includegraphics[clip, width=0.655\columnwidth]{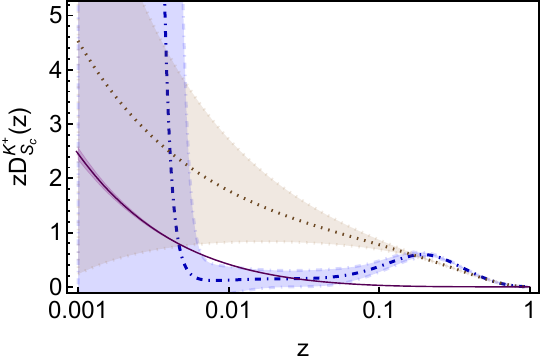} &
\includegraphics[clip, width=0.655\columnwidth]{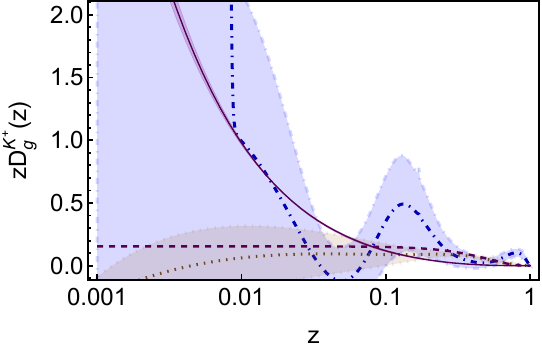} & \\
\end{tabular}
\caption{\label{FCSMJetK}
CSM results for kaon fragmentation functions, defined in Eqs.\,\eqref{SK}\,--\,\eqref{NKs}.
Solutions of cascade equations, Eq.\,\eqref{JetExplicit} -- dashed purple curves.
AO evolution of those curves to $\zeta=\zeta_2 := 2\,$GeV -- solid purple curves, with uncertainty bands obtained as described in Sect.\,\ref{HSuncertainty}.
Comparison curves are inferences from:
high-energy lepton-lepton, lepton-hadron and hadron-hadron scattering data \cite[JAM]{Moffat:2021dji} -- dotted brown curves, within like coloured bands;
and electron-positron annihilation and lepton-nucleon semi-inclusive deep-inelastic scattering data \cite[MAPFF]{AbdulKhalek:2022laj} -- dot-dashed blue curves within like-coloured bands.
}
\end{figure*}

\section{Realistic Fragmentation Functions}
\label{SecCSMFFs}
\subsection{CSM predictions}
\label{SubSecCSM}
Hadron scale pion and kaon dressed valence quark DFs were delivered in Ref.\,\cite{Cui:2020tdf}.  Drawn in Fig.\,\ref{CSMpiKDFs}\,A, they may be represented by the following functions:
{\allowdisplaybreaks
\begin{subequations}
\label{piKDFs}
\begin{align}
{\mathpzc u}_\pi(x;\zeta_{\cal H}) & =
{\mathpzc n}_\pi \ln \left[1 + \tfrac{1}{\rho_\pi^2} x^2(1-x)^2 \right.\nonumber \\
& \qquad \times \left. (1+ \tfrac{1}{2} \gamma_\pi^2 [([1-x]^2)^{\beta_\pi} +(x^2)^{\beta_\pi}])
\rule{0em}{2.2ex}\right]\,,
\end{align}
${\mathpzc n}_\pi = 0.858$,
$\rho_\pi = 0.116$,
$\gamma_\pi = 1.967$,
$\beta_\pi = 5.938$;
and
\begin{align}
{\mathpzc u}_K(x;\zeta_{\cal H}) & =
{\mathpzc n}_K \ln \left[ 1+\tfrac{1}{\rho_K^2} x^2 (1-x)^2 \right. \nonumber \\
& \qquad \times \left.  (1 + \gamma_K^2 (x^2)^{\alpha_K} ([1-x]^2)^{\beta_K}) \rule{0em}{2.3ex}\right]\,,
\end{align}
\end{subequations}
${\mathpzc n}_K = 0.444$,
$\rho_K = 0.0746$,
$\gamma_K = 6.276$,
$\alpha_K = 0.710$,
$\beta_\pi = 1.650$.
Again, ${\mathpzc s}_{K^-}(x;\zeta_{\cal H}) = {\mathpzc u}_{K}(1-x;\zeta_{\cal H})$.

Compared with the pointwise forms written in \linebreak Ref.\,\cite{Cui:2020tdf}, the functions in Eq.\,\eqref{piKDFs} are indistinguishable within visible line widths.  Stated mathematically, according to the standard ${\cal L}_1$ measure, the pion curves differ by 0.3\% and the kaon curves by $0.9$\%.  These differences are far smaller than the uncertainties associated with the original determinations; so, the new forms are equivalent by any reasonable assessment.
}

Regarding the DFs in Fig.\,\ref{CSMpiKDFs}\,A, it is worth reiterating some standard observations.
Namely, owing to EHM, both the pion and kaon DFs are significantly dilated with respect to the scale-free DF:
\begin{equation}
{\mathpzc q}_{\rm sf} = 30 x^2 (1-x)^2\,.
\label{ScaleFree}
\end{equation}
In addition, the kaon DFs are skewed as a consequence of Higgs boson (HB) couplings into QCD, which make the $s$ quark current mass 27-times larger than the mean light-quark mass \cite{ParticleDataGroup:2024cfk}.
The size of the skewing is suppressed by the magnitude of EHM, with the location of the peaks in the kaon DFs being shifted just $\pm 19$\% away from that in the pion DF.  This is commensurate with the scale set by $|1-f_\pi^2/f_K^2|$, \emph{viz}.\ by HB modulation of EHM as expressed in pseudoscalar meson leptonic decay constants.

Realistic pion and kaon EFFs are obtained from the DFs in Eq.\,\eqref{piKDFs} via the DLY relation, Eq.\,\eqref{DLYR}:
\begin{subequations}
\label{CSMEFFs}
\begin{align}
d_u^{\pi^+}(z) &= z {\mathpzc u}_\pi(1/z;\zeta_{\cal H}) \,, \\
d_u^{K^+}(z) & = z {\mathpzc u}_{K^+}(1/z;\zeta_{\cal H}) \,, \\
d_s^{K^-}(z) & = z {\mathpzc s}_{K^-}(1/z;\zeta_{\cal H}) \,.
\end{align}
\end{subequations}
They are drawn in Fig.\,\ref{CSMpiKDFs}\,B.
For comparison, the scale-free DF may be associated with the following EFF:
\begin{equation}
d^{\rm sf}(z) = \delta(z)\,.
\end{equation}
Normalising the EFFs according to Eq.\,\eqref{NormEFF}, then one finds the following CSM elementary $u$ quark multiplicities:
{\allowdisplaybreaks
\begin{subequations}
\begin{align}
m_u^\pi & \stackrel{\rm CSM}{=} \int_0^1 dz\, \tfrac{3}{2} d_{\mathpzc u}^{\pi^+}(z;\zeta_{\cal H}) = 0.80 \,, \\
m_u^K & \stackrel{\rm CSM}{=} \int_0^1 dz\, d_{\mathpzc u}^{K^+}(z;\zeta_{\cal H}) = 0.20 \,.
\end{align}
\end{subequations}
In matching the SCI values, Eq.\,\eqref{SCIEFFmq}, one is certain to obtain semiquantitative similarities between many SCI and CSM predictions.
Equation~\eqref{NormsKEFF} normalises $d_s^{K^-}(z) $.
}

Solving the hadron jet equations using the CSM EFFs defined by Eqs.\,\eqref{piKDFs}, \eqref{CSMEFFs}, one obtains the dashed purple curves in Figs.\,\ref{FCSMJetpi}\,A-C and  Figs.\,\ref{FCSMJetK}\,A-E.
Evolving those results by employing the procedure explained in Sect.\,\ref{FFEvolution}, including $s$ and $c$ quark mass thresholds -- see, \emph{e.g}., Ref.\,\cite[Sec.\,2]{Lu:2022cjx},
and setting
$\delta=0.11$;
one obtains the CSM predictions for $D_{S_{\mathpzc q},N_{\mathpzc q}}^h(z;\zeta_2=2\,{\rm GeV})$ drawn in Figs.\,\ref{FCSMJetpi}, \ref{FCSMJetK} (solid purple curves).

\subsection{Understanding SCI vs.\ CSM comparisons}
\label{SCIcfCSM}
As stressed in Sect.\,\ref{sectEFFs}, one should not typically view comparisons between SCI and CSM results as providing any sort of uncertainty estimate for our realistic CSM predictions.  It is nevertheless worth noting that SCI results are qualitatively and semiquantitatively in agreement with CSM predictions and this highlights the often cited utility of SCI analyses: they combine algebraic simplicity with a fair description of physical quantities.
In turn, this means there is value in identifying and explaining some of the more important differences between SCI results and CSM predictions.

\begin{table}[t]
\caption{\label{MomFracsCSM}
CSM fragmentation function momentum fractions obtained from solutions of the jet cascade equations at the hadron scale and after evolution to $\zeta_2$, following the scheme described in Sect.\,\ref{FFEvolution}.  As found using the SCI, $c\to q \to h$ contributions are negligible in all cases.
 }
\begin{center}
\begin{tabular*}
{\hsize}
{
l@{\extracolsep{0ptplus1fil}}
|l@{\extracolsep{0ptplus1fil}}
l@{\extracolsep{0ptplus1fil}}
|l@{\extracolsep{0ptplus1fil}}
l@{\extracolsep{0ptplus1fil}}
l@{\extracolsep{0ptplus1fil}}|}\hline
$h\ $ & \multicolumn{2}{c|}{$\pi^+ + \pi^0+\pi^-$} & \multicolumn{2}{c}{$K^+$} \\ \hline
 & $\zeta_{\cal H}\ $ & $\zeta_{2}\ $ & $\zeta_{\cal H}\ $ & $\zeta_2\ $ \\
$\langle z\rangle_{D_{S_{\mathpzc u}^{\mathpzc u}}}^h\ $ & $0.639\ $  & $0.418\ $  & $0.160\ $  & $0.105\ $ \\
$\langle z\rangle_{D_{S_{\mathpzc u}^{\mathpzc d}}}^h\ $ &  & $0.111\ $  &  & $0.028\ $ \\
$\langle z\rangle_{D_{S_{\mathpzc u}^{\mathpzc s}}}^h\ $ &  & $0.082\ $  &  & $0.021\ $ \\
$\langle z\rangle_{D_{S_{\mathpzc u}^{\mathpzc c}}}^h\ $ &  & $0.030\ $  &  & $0.008\ $ \\
$\langle z\rangle_{D_{S_{\mathpzc d}^{\mathpzc u}}}^h\ $ &   & $0.111\ $  &  & $0.016\ $ \\
$\langle z\rangle_{D_{S_{\mathpzc d}^{\mathpzc d}}}^h\ $ & $0.639\ $  & $0.418\ $  & $0.090\ $ & $0.059\ $ \\
$\langle z\rangle_{D_{S_{\mathpzc d}^{\mathpzc s}}}^h\ $ & & $0.082\ $  & & $0.011\ $ \\
$\langle z\rangle_{D_{S_{\mathpzc d}^{\mathpzc c}}}^h\ $ & & $0.030\ $  & & $0.004\ $ \\
$\langle z\rangle_{D_{S_{\mathpzc s}^{\mathpzc u}}}^h\ $ & & $0.057\ $  & & $0.049\ $ \\
$\langle z\rangle_{D_{S_{\mathpzc s}^{\mathpzc d}}}^h\ $ & & $0.057\ $  & & $0.049\ $ \\
$\langle z\rangle_{D_{S_{\mathpzc s}^{\mathpzc s}}}^h\ $ & $0.327\ $  & $0.199\ $  & $0.281\ $ & $0.171\ $ \\
$\langle z\rangle_{D_{S_{\mathpzc s}^{\mathpzc c}}}^h\ $ & & $0.016\ $  & & $0.013\ $ \\
%
$\langle z\rangle_{D_{{\mathpzc g}_{\mathpzc u}}^h}^\zeta\ $ & $0.080\ $  & 0.080 & $0.020\ $ & 0.020 \\
$\langle z\rangle_{D_{{\mathpzc g}_{\mathpzc d}}^h}^\zeta\ $ & $0.080\ $  & 0.080 & $0.011\ $ & 0.011 \\
$\langle z\rangle_{D_{{\mathpzc g}_{\mathpzc s}}^h}^\zeta\ $ & $0.041\ $  & $0.041$ & $0.035\ $  & $0.035\ $ \\
%
%
%
\hline
\end{tabular*}
\end{center}
\end{table}

Table~\ref{MomFracsCSM} -- see page~\pageref{MomFracsCSM} -- lists the parton species FF momentum fraction decompositions determined using the CSM EFFs.  Compared with the SCI analogue, Table~\ref{MomFracs}, kindred entries agree semiquantitatively in almost every case.
The exceptions are $\langle z\rangle_{D_{S_{\mathpzc s}^{\mathpzc s}}}^{\pi,K}$ and they are readily understood.

As evidenced by Eq.\,\eqref{FFstopi}, the $s\to \pi$ FF is unfavoured.
It proceeds via the convolutions \linebreak
$d_{\mathpzc s}^{\mathpzc u}(\frac{z}{y})  \otimes D_{\mathpzc u}^{\pi^+}(y)$ and
$d_{\mathpzc s}^{\mathpzc d}(\frac{z}{y})  \otimes D_{\mathpzc d}^{\pi^+}(y)$.
The FFs $D_{{\mathpzc u},{\mathpzc d}}^{\pi}(z)$ are favoured, so both SCI and CSM results possess strong support on the entire $z$ domain; see Figs.\,\ref{FSCIJetpi}\,B, \ref{FCSMJetpi}\,B.
On the other hand, the SCI and CSM EFFs are very different: whereas the CSM forms are roughly symmetric around $z=1/2$ and endpoint suppressed -- see Fig.\,\ref{CSMpiKDFs}\,B; using the SCI,
$d_{\mathpzc s}^{\mathpzc u}(z) = d_{\mathpzc s}^{\mathpzc d}(z)
= d_s^{\bar K^0}(1-z)$ is asymmetric, enhanced on $z\lesssim 0.5$ and strongly damping on $z\gtrsim 0.5$ -- see Fig.\,\ref{SCIpiKDFs}\,B.
Consequently, solving the convolution cascade equations yields an SCI result for $D_{S_{\mathpzc s}^{\mathpzc s}}^{\pi}$ that is quite unlike that obtained using the CSM inputs.
Indeed, comparing Fig.\,\ref{FCSMJetpi}\,C with Fig.\,\ref{FSCIJetpi}\,C, one sees that the former is larger in magnitude and possesses a domain of strong support that stretches closer to $z\simeq 1$.
Hence, it delivers a significantly larger momentum fraction.

Considering $\langle z\rangle_{D_{S_{\mathpzc s}^{\mathpzc s}}}^{K}$, the FF
$D_{S_{\mathpzc s}^{\mathpzc s}}^{K}$ is favoured -- see Eq.\,\eqref{FFstoKm}, in which the EFF driving term is $d_{\mathpzc s}^{K^-}(z)$.
The SCI result for this function is strongly enhanced on $z\gtrsim 0.5$ -- Fig.\,\ref{SCIpiKDFs}\,B, whereas the CSM form falls toward zero on that domain -- Fig.\,\ref{CSMpiKDFs}\,B.
Consequently, the SCI result for $D_{S_{\mathpzc s}^{\mathpzc s}}^{K}$ has stronger support at large $z$ than the CSM prediction and thus delivers a larger value for $\langle z\rangle_{D_{S_{\mathpzc s}^{\mathpzc s}}}^{K}$: Fig.\,\ref{FSCIJetK}\,E \emph{cf}.\ Fig.\,\ref{FCSMJetK}\,E.

Having understood the results in Table~\ref{MomFracsCSM}, they can now be used to demonstrate momentum conservation for all CSM FFs.  One need only replace the Table~\ref{MomFracs} entries in Eqs.\,\eqref{MomConsAllhu}\,--\,\eqref{MomConsAllhs} with their Table~\ref{MomFracsCSM} analogues.


\subsection{Comparison with phenomenological inferences}
\label{SecCompare}
As above, for comparison with our predictions, the inferences from data reported in Refs.\,\cite{Moffat:2021dji, AbdulKhalek:2022laj} are also drawn in Figs.\,\ref{FCSMJetpi}, \ref{FCSMJetK}.
We have already noted that the fits are mutually incompatible on $z\lesssim 0.5$.
Compared with our predictions, the situation is equally poor; namely, there is little agreement.
\smallskip

\noindent First consider the pion.
\begin{description}
  \item[Figs.\,\ref{FCSMJetpi}\,A, B.] $u \to \pi$ (favoured), nonsinglet and singlet.  There is agreement only on $z\gtrsim 0.5$, \emph{i.e}., on the valence quark domain.
      Further, the JAM nonsinglet FF result ($z D_N$) exhibits an unexpected divergence on $z\simeq 0$.  This is the domain of glue and sea dominance; so given Eq.\,\eqref{Npi}, $z D_N$ should vanish.
  \item[Fig.\,\ref{FCSMJetpi}\,C.] $s\to \pi$.  One might say that there is qualitative agreement on the far valence domain, but only in the sense that this FF is small.  Otherwise, any agreement is only the result of an accidental curve crossing.
  \item[Fig.\,\ref{FCSMJetpi}\,D-E.] $c,g\to \pi$.  Plainly, there is no agreement on these FFs, which are very poorly constrained by data.  
      Our predictions can potentially provide a coherent picture of fragmentation across all parton species.
\end{description}

\noindent Now turn to the kaon.
\begin{description}
  \item[Figs.\,\ref{FCSMJetK}\,A, B.]  $u \to K$  (favoured), nonsinglet and singlet.  Similar to the pion solutions, there is agreement only on $z\gtrsim 0.4$.
      Here, the JAM result for $z D_N$ is finite and nonzero on $z\simeq 0$, which is again unexpected.
      Moreover, $z D_S$ is also nonzero and finite, in contradiction of the analogous $u\to \pi$ result and our prediction.
  \item[Figs.\,\ref{FCSMJetK}\,D, E] $s\to K$ (favoured), nonsinglet and singlet.  Agreement is seen on $z\gtrsim 0.7$; but nothing beyond that.  Both JAM and MAPFF produce nonzero finite values on $z\simeq 0$, where, on physics grounds, such outcomes are not expected.
  \item[Figs.\,\ref{FCSMJetK}\,C,] $d \to K$.  One might claim qualitative agree\-ment on the far valence domain, but again only because this FF is small.  Furthermore and once more unexpectedly, JAM and MAPFF fits produce non\-zero finite values on $z\simeq 0$.  Naturally, our predictions diverge on this glue and sea dominated domain.
  \item[Figs.\,\ref{FCSMJetK}\,F, G.] $c, g \to K$.  Again, there is no agreement on these FFs, which are very poorly constrained by data; 
      and our predictions can potentially provide a coherent picture of fragmentation across all parton species.
\end{description}

\begin{figure}[t]
\leftline{\large\sf A}
\vspace*{-2ex}

\centerline{%
\includegraphics[clip, width=0.95\linewidth]{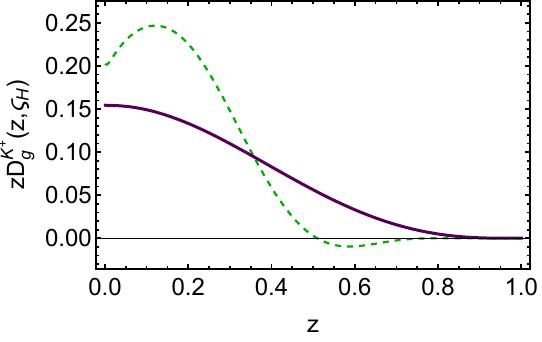}}

\vspace*{1ex}

\leftline{\large\sf B}
\vspace*{-2ex}

\centerline{%
\hspace*{-1ex}\includegraphics[clip, width=0.96\linewidth]{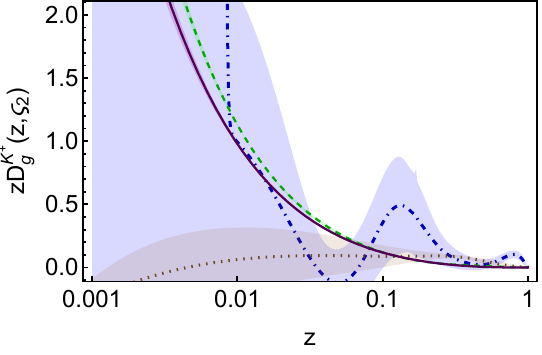}}
\caption{\label{FigGlueProf}
{\sf Panel A}.  Hadron scale in-kaon gluon FF profiles, Eq.\,\eqref{DgDSpi}: SCI -- dashed green curve; CSM -- solid purple.
{\sf Panel B}. In-kaon gluon FF profiles at $\zeta=\zeta_2 := 2\,$GeV: SCI -- dashed green curve; CSM -- solid purple.
Comparison curves are inferences from:
high-energy lepton-lepton, lepton-hadron and hadron-hadron scattering data \cite[JAM]{Moffat:2021dji} -- dotted brown curves, within like coloured bands;
and electron-positron annihilation and lepton-nucleon semi-inclusive deep-inelastic scattering data \cite[MAPFF]{AbdulKhalek:2022laj} -- dot-dashed blue curves within like-coloured bands.
%
%
}
\end{figure}

It is worth highlighting that the non-monotonic (oscillatory) behaviour of the MAPFF fits on $z\lesssim 0.5$ is entirely incompatible with our predictions.  Indeed, quite generally, the MAPFF results suggest strongly that FFs are practically unconstrained on $z\lesssim 0.2$.  The observations and remarks collected here indicate that, today, phenomenology does not deliver objective FF results: the results obtained are practitioner specific.

In completing this comparison, we return to Sect.\,\ref{glueFFSuncertainty} and address the issue of whether an uncertainty in the gluon FF profile, Eq.\,\eqref{DgDSpi}, has an impact on the preceding discussion.  In this connection, consider Fig.\,\ref{FigGlueProf}.
Panel A compares hadron-scale SCI and CSM \emph{Ans\"atze} for $D_g^{K^+}$: each is an interaction-consistent profile; and they display qualitative similarities, but significant quantitative differences.
Turn now to Panel B, which compares the $\zeta_{\cal H}\to \zeta_2$ evolved FFs: SCI with CSM and both with  the fits to data in Refs.\,\cite[JAM]{Moffat:2021dji}.  
Plainly, estimated as we consider reasonable -- Sect.\,\ref{glueFFSuncertainty} -- any systematic uncertainty in our prediction for $D_g^{K^+}(z;\zeta_2)$, in particular and, likewise all others, is negligible in comparison with those that characterise existing FF fits.
It is therefore practically immaterial.

\section{Hadron Jet Multiplicities}
\label{HadronMultiplicities}
Working from Ref.\,\cite[Eq.\,(3.2)]{Field:1976ve}, the quantity
\begin{equation}
    M_{\mathpzc p}^h(\zeta) = \int_{z_{\rm min}}^1 dz\, D_{\mathpzc p}^h(z;\zeta)
    \label{FFmultzmin}
\end{equation}
is the mean multiplicity of hadrons, $h$, emerging from the parent parton $p$ with $z >z_{\rm min}$.  Since, as we have seen, $\zeta>\zeta_{\cal H}$ FFs diverge faster than $1/z$ on $z\simeq 0$, then $M_{\mathpzc p}^h$ increases without bound as the momentum of the parent parton increases.

{\allowdisplaybreaks
In the context of realisable experiments, consider $e^+ e^- \to h X$.  An associated multiplicity structure function is normally defined as follows \cite[Sec.\,3.1.1]{Metz:2016swz}:
\begin{equation}
    F^h(z;\zeta)
    = \frac{1}{\sigma_{\rm tot}}\frac{d\sigma^{e^+e^- \to h X}}{dz}\,,
\end{equation}
with $\sigma_{\rm tot}$ being the total cross-section, so that $F^h(z;\zeta)$ is the number of $h$ hadrons produced in each event.
%
At leasing order in perturbative QCD, 
\begin{equation}
\frac{d\sigma^{e^+e^- \to h X}}{dz}
= \sum_{\mathpzc q}e_{\mathpzc q}^2 D_{\mathpzc q}^h(z;\zeta)\,,
\end{equation}
and $\sigma_{\rm tot} = \sum_{q}e_q^2 =: \sigma_{\rm tot}^{\mbox{\rm\tiny  LO}}$.  
In terms of the multiplicity structure function, the total multiplicity is:
\begin{equation}
M^h(\zeta) = \sum_{\mathpzc p} \int_{z_{\rm min}}^1 dz\, F_{\mathpzc p}^h(z;\zeta)\,.
\end{equation}
}

Conversion between experimental kinematics and $z$ is typically achieved by defining
\begin{equation}
    z = 2  E_h/\sqrt{Q^2}\,,
    \label{zExp}
\end{equation}
where $Q^2=\zeta^2$ is the momentum transfer provided by the $e^+ e^-$ collision.
Using Eq.\,\eqref{zExp}, it is clear that the minimum available value of the fragmentation momentum fraction is
\begin{equation}
    z_{\rm min} = 2\,{\rm mass}_{\rm produced\;hadron}/\zeta\,.
    \label{zmin}
\end{equation}
Namely, the mass of the produced hadron places a natural lower bound on the integral in Eq.\,\eqref{FFmultzmin}.
Evidently, in line with the statements made above, $M_{\mathpzc p}^h$ grows with increasing $\zeta$.

In reality, $\sigma_{\rm tot}$ can only be calculated nonperturbatively and the measured value depends on many things, including experimental setup, detector acceptance, etc.  Thus, in subsequent comparisons with experimental results for the $z$-dependent multiplicity distribution, we write 
\begin{equation}
F^h(z;\zeta) = \frac{1}{{\cal N}}
\sum_{\mathpzc q}e_{\mathpzc q}^2 D_{\mathpzc q}^h(z;\zeta)\,,
\label{FhNorm}
\end{equation}
with ${\cal N}$ chosen so as to ensure a fair match between our prediction and experiment for the integrated multiplicity over the central domain $z\in[0.1,0.7]$.  We choose this domain because, in our view, the experimental uncertainty in extant experimental results on the complement of this domain, \emph{i.e}., the deep sea and far valence domains, are underestimated.

Owing to ${\cal G}$-parity symmetry, one may reliably obtain the total pion multiplicity from a measured charged-pion value using the formula:
\begin{equation}
M^\pi(\zeta) = \tfrac{3}{2} [ M^{\pi^+} (\zeta)+ M^{\pi^-} (\zeta)] \,.
\label{PionRatio0}
\end{equation}
This raises the following question: Given a measured charged kaon multiplicity, is there an analogous formula by which one can estimate the total kaon multiplicity?

Supposing SU$(3)$-flavour symmetry were exact, then one would have
\begin{equation}
M^K(\zeta)  \approx 2 [ M^{K^+} (\zeta)+ M^{K^-} (\zeta)] \,,
\label{KaonRatio0}
\end{equation}
which is a statement of the assumption:
$M^{K^+} (\zeta)+ M^{K^-} (\zeta) \approx M^{K^0} (\zeta)+ M^{\bar K^0} (\zeta)$.  However, SU$(3)$-flavour symmetry is not exact; so, it is desirable to estimate the correction to Eq.\,\eqref{KaonRatio0}.

{\allowdisplaybreaks
To proceed, therefore, using both SCI and CSM FFs, we computed the charged/neut\-ral multiplicity ratio:
\begin{equation}
\label{RKzeta0}
R_K(\zeta) = \frac{M^{K^+} (\zeta)+ M^{K^-} (\zeta)}{M^{K^0} (\zeta)+ M^{\bar K^0} (\zeta)}\,.
\end{equation}
In detail, using the identities and relations above, one finds the following expressions for the numerator and denominator:
\begin{subequations}
\label{RKzeta}
\begin{align}
M^{K^+}& (\zeta) + M^{K^-} (\zeta) =
\int_{z_{\rm min}}^1 dz\, \big[4 D_{S_{\mathpzc u}}^{K^+}(z;\zeta) \nonumber \\
&
+D_{S_{\mathpzc d}}^{K^+}(z;\zeta)+D_{S_{\mathpzc s}}^{K^+}(z;\zeta) + 4 D_{S_{\mathpzc c}}^{K^+}(z;\zeta)\big] \,, \label{RKzetaN}\\
M^{K^0}& (\zeta)  + M^{\bar K^0} (\zeta) =
\int_{z_{\rm min}}^1 dz\, \big[4 D_{S_{\mathpzc d}}^{K^+}(z;\zeta) \nonumber \\
&
+D_{S_{\mathpzc u}}^{K^+}(z;\zeta)+D_{S_{\mathpzc s}}^{K^+}(z;\zeta) + 4 D_{S_{\mathpzc c}}^{K^+}(z;\zeta)\big] \,, \label{RKzetaD}
\end{align}
\end{subequations}
where, naturally, Eq.\,\eqref{RKzetaN} maps into Eq.\,\eqref{RKzetaD} under $u \leftrightarrow d$.
Since
\begin{equation}
D_{S_{\mathpzc u}}^{K^+}(x;\zeta_2) \neq D_{S_{\mathpzc d}}^{K^+}(x;\zeta_2) = D_{S_{\mathpzc u}}^{K^0}(x;\zeta_2)\,,
\end{equation}
see Fig.\,\ref{FSCIJetK}\,B \emph{cf}.\ Fig.\,\ref{FSCIJetK}\,C and Fig.\,\ref{FCSMJetK}\,B \emph{cf}.\ Fig.\,\ref{FCSMJetK}\,C, then
\begin{equation}
R_K(\zeta_2) \neq 1\,.
\end{equation}
}

\begin{figure}[t]
\centerline{%
\includegraphics[clip, width=0.95\linewidth]{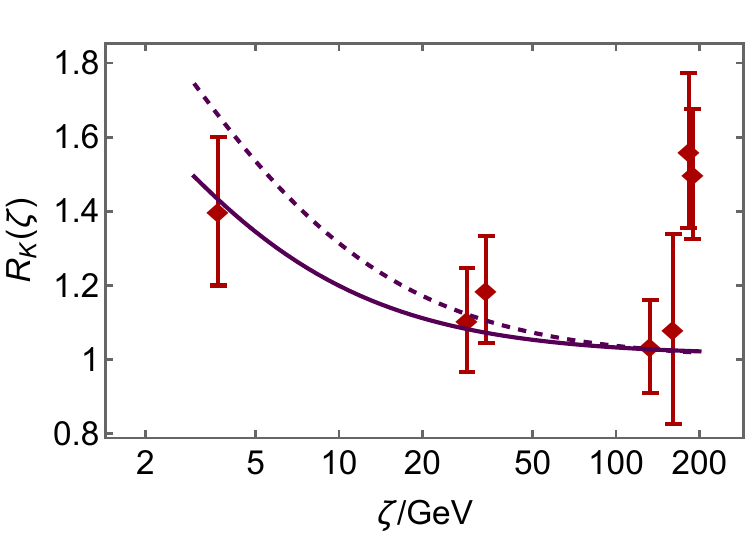}}
\caption{\label{FigRK}
SCI and CSM predictions for the $\zeta$-dependence of the relative multiplicity of charged and neutral kaons, Eq.\,\eqref{RKzeta0}, \eqref{RKzeta}.
Data are empirical estimates from Refs.\,\cite{BESIII:2025mbc, TPCTwoGamma:1983lrv, TPCTwoGamma:1984eoj, TASSO:1988jma, DELPHI:2000ahn}.
See also Table~\ref{TableRK}.
}
\end{figure}

On the other hand, as $\zeta$ increases, FF support is transferred to the domain of glue and sea dominance, whereupon valence-quark induced differences are in\-crea\-sin\-gly suppressed.
Consequently, $R_K(\zeta)$ must decrease toward unity with increasing $\zeta$.
This is evident from the SCI and CSM results reported in Table~\ref{TableRK} and displayed in Fig.\,\ref{FigRK}.

\begin{table}[t]
\caption{\label{TableRK}
SCI and CSM predictions for the $\zeta$-dependence of the relative multiplicity of charged and neutral kaons, Eq.\,\eqref{RKzeta0}, \eqref{RKzeta}.
Also listed are empirical estimates from Refs.\,\cite{BESIII:2025mbc, TPCTwoGamma:1983lrv, TPCTwoGamma:1984eoj, TASSO:1988jma, DELPHI:2000ahn}.
(Dimensioned quantities in GeV.)
}
\begin{center}
\begin{tabular*}
{\hsize}
{
l@{\extracolsep{0ptplus1fil}}
|l@{\extracolsep{0ptplus1fil}}
l@{\extracolsep{0ptplus1fil}}}\hline
Predictions & $\zeta\ $ & $R_K\ $  \\\hline
SCI & $\phantom{11}3.05\ $ & $1.73\ $ \\
 & $\phantom{11}3.67\ $ & $1.67\ $\\
 & $\phantom{1}10\phantom{.00}\ $ & $1.31\ $\\
 & $\phantom{1}91.2\ $ & $1.038\ $\\
 & $189\phantom{.00}\ $ & $1.022\ $\\
\hline
Predictions & $\zeta\ $ & $R_K\ $  \\\hline
CSM & $\phantom{11}3.05\ $ & $1.49\ $ \\
 & $\phantom{11}3.67\ $ & $1.43\ $\\
 & $\phantom{1}10\phantom{.00}\ $ & $1.20\ $\\
 & $\phantom{1}91.2\ $ & $1.035\ $\\
 & $189\phantom{.00}\ $ & $1.022\ $\\
\hline
Measurements & $\zeta\ $ & $R_K\ $  \\\hline
\cite[BESIII]{BESIII:2025mbc} & $\phantom{11}3.67\ $  & 1.40(20) \\
\cite[TPC]{TPCTwoGamma:1983lrv, TPCTwoGamma:1984eoj}$\ $ & $\phantom{1}29\phantom{.00}\ $ & $1.11(16)\ $\\
\cite[TASSO]{TASSO:1988jma}$\ $ & $\phantom{1}34\phantom{.00}\ $ & $1.19(14)\ $\\
\cite[DELPHI]{DELPHI:2000ahn} $\ $ & $133\ $ & $1.04(13)\ $ \\
                                                   & $161\ $ & $1.08(26)\ $ \\
                                                   & $183\ $ & $1.56(21)\ $ \\
                                                   & $189\ $ & $1.50(18)\ $ \\
\hline
\end{tabular*}
\end{center}
\end{table}

Some available empirical information on $R_K(\zeta)$ is also presented in Table~\ref{TableRK} and  Fig.\,\ref{FigRK}.
The SCI and CSM $\zeta$-trajectories are qualitatively confirmed by the data, with the CSM prediction delivering the better quantitative agreement.
It is worth noting that the large $\zeta$ data in Ref.\,\cite[DELPHI]{DELPHI:2000ahn} are only marginally consistent internally: the two points at largest $\zeta$ sit unexpectedly high.
Further, Ref.\,\cite{BESIII:2025mbc} does not report an uncertainty.  For illustrative purposes, therefore, we have drawn an error on this datum that is determined by the mean relative uncertainty of the other data.
Finally, an excess of charged vs.\ neutral kaons is also observed in high-energy collisions of atomic nuclei ($\surd s = 11.9\,$GeV) \cite{NA61SHINE:2023azp} and discussed in Ref.\,\cite{Reichert:2025znn}.

Hereafter, in order to obtain total kaon multiplicities from the charged kaon value, we employ the CSM result for $R_K(\zeta)$:
\begin{equation}
M^K(\zeta)  = [1 + 1/R_K^{\rm CSM}(\zeta)] [ M^{K^+} (\zeta)+ M^{K^-} (\zeta)] \,.
\label{KaonRatio}
\end{equation}
As Fig.\,\ref{FigRK} elucidates, the correction is only important on $\zeta/m_p \lesssim 50$.

Following this preparation, consider now the experiments described in Refs.\,\cite{OPAL:1994zan, DELPHI:1998cgx, SLD:2003ogn, CLEO:1984ylq, ARGUS:1989zdf, BaBar:2013yrg}, results from which may be viewed as delivering $\pi, K$ multiplicities in $e^+e^-$ \linebreak $\to h X$ at two different energies $\sqrt s = m_{Z^0} = 91.2\,$GeV \cite{OPAL:1994zan, DELPHI:1998cgx, SLD:2003ogn} and $\sqrt s \approx 10\,$GeV \cite{CLEO:1984ylq, ARGUS:1989zdf, BaBar:2013yrg}.  The results are reported in Table~\ref{TableMult}, wherein, for the experiments, charged particle multiplicities are converted to total multiplicities using Eqs.\,\eqref{PionRatio0}, \eqref{KaonRatio}.
There is one possible exception.  Namely, Ref.\,\cite[CLEO]{CLEO:1984ylq} included some neutral kaons in their total kaon yield; so, both converted (C) and unconverted (U) results are listed in Table~\ref{TableMult}.

\begin{table}[t]
\caption{\label{TableMult}
Fractional $\pi, K$ multiplicities in $e^+ e^- \to h X$ from Refs.\,\cite{OPAL:1994zan, DELPHI:1998cgx, SLD:2003ogn, CLEO:1984ylq, ARGUS:1989zdf, BaBar:2013yrg} compared with SCI and CSM predictions.
(Dimensioned quantities in GeV.  $p, \bar p$ production is neglected because the multiplicities are typically $\lesssim 3$\%.)
}
\begin{center}
\begin{tabular*}
{\hsize}
{
l@{\extracolsep{0ptplus1fil}}
l@{\extracolsep{0ptplus1fil}}
|l@{\extracolsep{0ptplus1fil}}
l@{\extracolsep{0ptplus1fil}}}\hline
Measurements & $\sqrt s\ $ & $\pi\ $ & $K\ $ \\\hline
\cite[OPAL]{OPAL:1994zan} & $91.2\ $ & $0.84(1)\ $& $0.16(1)\ $\\
\cite[DELPHI]{DELPHI:1998cgx} & $91.2\ $ & $0.86(6)\ $ & $0.14(1)\ $\\
\cite[SLD]{SLD:2003ogn} & $91.2\ $ & $0.86(1)\ $& $0.14(1)\ $\\
\cite[CLEO]{CLEO:1984ylq}$_{\rm C}\ $ & $10.5\ $ & $0.84(6)\ $ & $0.16(3)\ $ \\
\cite[CLEO]{CLEO:1984ylq}$_{\rm U}\ $ & $10.5\ $ & $0.91(6)\ $ & $0.09(2)\ $ \\
\cite[ARGUS]{ARGUS:1989zdf} & $10.0\ $ & $0.85(2)\ $ & $0.15(1)\ $\\
\cite[BaBar]{BaBar:2013yrg} & $10.5\ $ & $0.85(3)\ $ & $0.15(1)\ $\\
\hline
Predictions & $\sqrt s\ $ & $\pi\ $ & $K\ $ \\\hline
SCI & $91.2\ $ & $0.81\ $ & $0.19\ $ \\
CSM & $91.2\ $ & $0.83\ $ & $0.17\ $ \\
SCI & $10.0\ $ & $0.85\ $ & $0.15\ $ \\
CSM & $10.0\ $ & $0.87\ $ & $0.13\ $ \\
\hline
\end{tabular*}
\end{center}
\end{table}

\begin{figure}[t]
\centerline{%
\includegraphics[clip, width=0.95\linewidth]{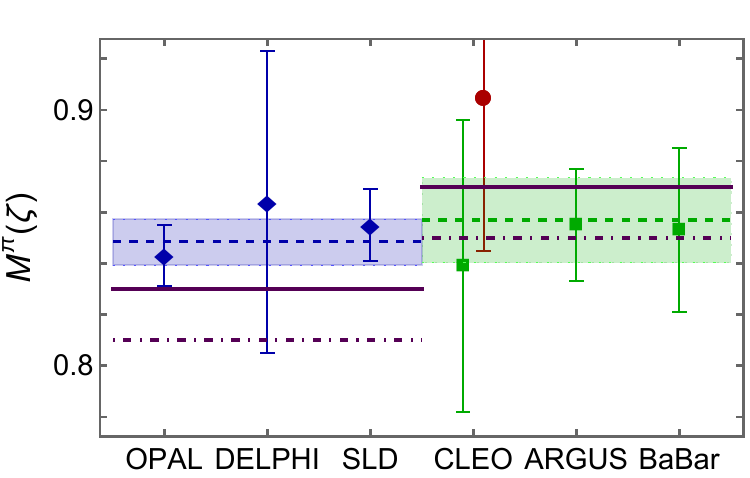}}
\caption{\label{FigMultPion}
Points: fractional pion multiplicities listed in Table~\ref{TableMult}.  The dashed lines within like coloured bands are the uncertainty weighted averages in Eq.\,\eqref{UWMean}.
CSM (solid purple) and SCI (dot-dashed purple) predictions at each energy, reproduced from Table~\ref{TableMult}.
(Kaon results are obtained by number conservation: the total must be unity.)
}
\end{figure}

The experimental results for pion multiplicities in Table~\ref{TableMult} are drawn in Fig.\,\ref{FigMultPion}.  The dashed horizontal lines within like coloured bands are the uncertainty weighted averages at each energy:
\begin{equation}
M^\pi(91\,{\rm GeV}) = 0.848(09)\,,\;
M^\pi(10\,{\rm GeV})  = 0.857(17)\,.
\label{UWMean}
\end{equation}
Both CLEO results were used here; but since the uncertainties on these points are large, the impact is small -- the result changes by $<1$\% if the unconverted value is omitted.
The data hint at an energy dependence of the pion/ka\-on multiplicity ratio, with that ratio increasing as energy ($\surd s$) decreases.

Figure~\ref{FigMultPion} also includes CSM and SCI predictions for the energy-dependent multiplicity ratio.  Evidently, consistent with the data suggestion, theory predicts that the ratio increases as the energy is decreased.
This is a natural outcome because as energy increases, the hadron masses become irrelevant.
Consequently, with increasing energy, the pion/kaon multiplicity ratio should fall to meet some asymptotic value that is uninfluenced by the hadron mass threshold introduced by Eq.\,\eqref{zmin}, being instead determined solely by the FFs.  Similar behaviour is also typically found when using phenomenological fits.

\begin{figure}[t]
\leftline{\large\sf A}
\vspace*{-2ex}

\centerline{%
\includegraphics[clip, width=0.95\linewidth]{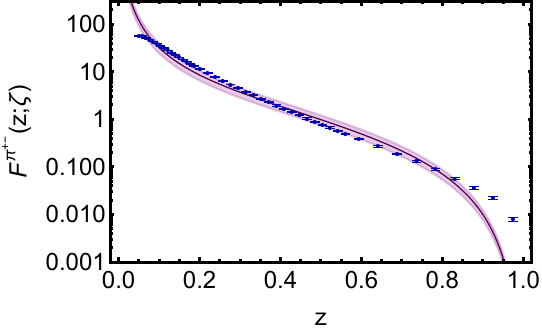}}

\vspace*{1ex}

\leftline{\large\sf B}
\vspace*{-2ex}

\centerline{%
\hspace*{-1ex}\includegraphics[clip, width=0.96\linewidth]{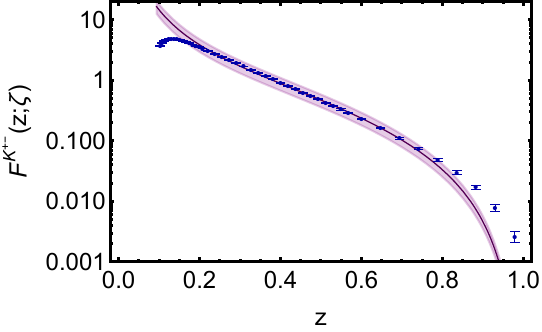}}
\caption{\label{zdepmult}
Hadron multiplicity structure function in Eq.\,\eqref{FhNorm} at $\zeta = 10.5\,$GeV -- solid purple curves with associated uncertainty bands -- compared with data drawn from Ref.\,\cite[BaBar]{BaBar:2013yrg}.
{\sf Panel A}.  $h = \pi^\pm$.
{\sf Panel B}.  $h = K^\pm$.
The theory uncertainty bands are explained in the text.
%
%
}
\end{figure}

Having thus far discussed integrated multiplicities, it is here worth illustrating the comparison between data and our predictions for the $z$-dependent multiplicities used to calculate the above results.  For this purpose, consider the most recent available $e^+ e^- \to h X$ data \cite{BaBar:2013yrg}.   Proceeding therewith, as discussed in connection with Eq.\,\eqref{FhNorm}, one obtains the pion and kaon results drawn in Fig.\,\ref{zdepmult}.
Regarding our predictions, the band expresses an uncertainty defined by the difference between ${\cal N}$ calculated using the $\pi^\pm$ or $K^\pm$ multiplicity -- see discussion of Eq.\,\eqref{FhNorm}: 
$1/{\cal N} = 4.0(1.0) /\sigma_{\rm tot}^{\mbox{\rm\tiny  LO}}$.
In both cases, the pointwise agreement is good on the domain $z\in [0.1,0.7]$.  
It follows, furthermore, that the $z$-dependent ratio of $\pi^\pm$ and $K^\pm$ multiplicity structure functions is also in agreement with data; and in this comparison, ${\cal N}$ plays no role.
(Comparisons for the other sets are similar, \emph{e.g}., working with the $\sqrt s = 29\,$GeV data in Ref.\,\cite[TPC]{TPCTwoGamma:1983lrv}, $1/{\cal N} = 3.5(7) \sigma_{\rm tot}^{\mbox{\rm\tiny  LO}}$ delivers pointwise agreement that is equal to or better than that displayed in Fig.\,\ref{zdepmult}.)

Referred to the CSM predictions in Fig.\,\ref{zdepmult}, the data appear to underestimate the hadron yield on the deep-sea domain and overestimate it at far-valence $z$. 
In our view, it is unsurprising that measurements underestimate the yield of low-$z$ hadrons.  In principle, as discussed after Eq.\,\eqref{FFmultzmin}, when neglecting hadron mass effects, infinitely many $z\simeq 0$ $h$-hadrons are produced from each parton.  This is evident in our evolved cascade FF predictions.  Detector efficiency and related experimental issues likely degrade acceptance on any neighbourhood of this domain.

Turning to the far-valence domain, the experimental particle yield is very small and rapidly decreasing on $z\gtrsim 0.7$, \emph{e.g}., the hadron number in the $z\approx 0.7$ bin is just 10\% of that in the $z\approx 0.5$ bin and the yield in the largest $z$-bin is less-than $0.4$\% of that in the $z\approx 0.5$ bin.
With such small yields, it is possible that the measurement's systematic error is underestimated at far-valence $z$.  We now discuss another possibility.  

We first stress that leading-order formulae have been used as the means of translating our FF predictions into multiplicities for comparisons with data.  They provide the simplest bridge.  
One might amend the approach by modifying the hard scattering kernel.  For instance, hard scattering kernels for use in making these connections are known in QCD perturbation theory to O$(\alpha_s^2)$ \cite{Rijken:1996ns}, where $\alpha_s$ is the strong running coupling.
However, regarding DFs, analyses have shown that increasing the $\alpha_s$-order of corrections included in the associated hard scattering kernel does not make a big difference.  The effect of such terms can largely be contained in a change of the matching/factorisation scale.  
Instead, the key appears to lie in going beyond leading logarithm resummation in the kernel; see, \emph{e.g}., Ref.\,\cite{Cui:2021mom}.  This has a very significant effect on the valence quark domain \cite[Fig.\,7A]{Cui:2021mom}.

Given the DLY relation, then the same is true for FFs; and this is, perhaps, what is being signalled by the discrepancy between our predictions for the $z$-dependent multiplicities and data on $z\gtrsim 0.7$: potentially, on this domain, NLL-improved hard scattering kernels should be used to build the bridge to data.
However, such kernels are not yet available.  

It is worth remarking that the analogous approach to connecting proton DFs with modern data on the structure function ratio $F_2^n/F_2^p$ \cite{JeffersonLabHallATritium:2021usd} has proven efficacious \cite[Fig.\,2A]{Chang:2022jri}.  This is additional circumstantial support for the approach we have adopted.

\section{Summary and Perspective}
\label{Epilogue}
Exploiting the Drell-Levy-Yan (DLY) relation \cite{Drell:1969jm, Drell:1969wd, Drell:1969wb, Gribov:1972ri, Gribov:1972rt}, in-had\-ron dressed-valence parton distribution functions (DFs) for the pion and kaon were used to define hadron-scale, $\zeta_{\cal H}<m_p$, parton-to-hadron elementary fragmentation functions (EFFs).  ($m_p$ is the proton mass.)
Two distinct source DF sets were used, \emph{viz}.\ one obtained using a symmetry-preserving treatment of a vector\,$\times$\,vec\-tor contact interaction (SCI) and the other representing available predictions delivered by continuum Schwinger function methods (CSMs) [Sects. \ref{SecEFFs}, \ref{SecCSMFFs}].
Using the EFFs thus obtained as the driving terms in a coupled set of hadron cascade equations [Sect.\,\ref{HJEs}], complete hadron-scale fragmentation functions (FFs) for pion and kaon production in high energy reactions were subsequently obtained [Sects.\,\ref{SecSCIFFs}, \ref{SecCSMFFs}].

The hadron-scale FFs were evolved to scales accessible in experiment using the all-orders scheme [Sect.\,\ref{FFEvolution}].
The evolution equations do not alone ensure momentum conservation for quark singlet FFs; but there is a value of the momentum fraction stored in gluon FFs such that momentum is conserved under evolution in the sum over all singlet FFs.  The same fraction ($\approx 11$\% for four quark flavours) ensures momentum conservation for any form of input FFs.

Compared with each other at the resolving scale $\zeta=\zeta_2$, SCI and CSM FF predictions are in qualitative and, typically, semiquantitative agreement [Figs.\,\ref{FSCIJetpi}, \ref{FSCIJetK}, \ref{FCSMJetpi}, \ref{FCSMJetK}].  Importantly, the predictions conform with all QCD-based expectations for behaviour on the endpoint domains $z\simeq 0, 1$, \emph{e.g}., nonsinglet FFs vanish at $z=0$ and singlet FFs diverge faster than $1/z$.  The quantitative disagreements between a few SCI and CSM FFs are understood as reflecting limitations of the SCI.

On the other hand, phenomenological inferences of FFs from data \cite{Hirai:2007cx, deFlorian:2014xna, Bertone:2017tyb, Soleymaninia:2020bsq, Moffat:2021dji, AbdulKhalek:2022laj, Gao:2024dbv} are
mutually inconsistent on $z\lesssim 0.5$ and often on a larger domain;
fail to conform with expected endpoint behaviour, \emph{e.g}., with singlet FFs that satisfy $z D_{\rm singlet}(z,\zeta_2)<\infty$ on $z\simeq 0$, whereupon glue and sea contributions should lead to divergences;
and largely incompatible with the predictions delivered here\-in [Sect. \ref{SecCompare}], hence unrelated to solutions of hadron cascade equations \cite{Field:1976ve, Field:1977fa, Altarelli:1981ax}.

Predictions for hadron jet multiplicities were also delivered [Sect.\,\ref{HadronMultiplicities}].  Since proton, antiproton yields are small ($\lesssim 3$\%), then, in comparison with data, $\pi, K$ yields were considered to be practically exhaustive.
The predictions reveal SU$(3)$ symmetry breaking in the char\-ged-kaon/neutral-kaon multiplicity ratio, which is significant at reaction energy scales $\zeta \approx 3 m_p$, but decreases in size with increasing reaction energy [Fig.\,\ref{FigRK}].
They also show that the pion/kaon ratio in $e^+ e^- \to h X$ is energy dependent: as $\zeta$ increases, the ratio diminishes to a value that is independent of hadron masses [Fig.\,\ref{FigMultPion}].
Moreover, CSM predictions for the individual $z$-dependent $\pi^\pm$, $K^\pm$ multiplicities were shown to be in good agreement with available data on $z\in[0.1,0.7]$ [Fig.\,\ref{zdepmult} and associated discussion].

The analysis herein suggests that CSM FF predictions should be seen as, at least, providing useful guidance for future data analyses and, in themselves,  potentially serving as realistic descriptions of hadronisation.
Regarding guidance, they give clear indications on the endpoint behaviour that should be expressed by realistic FFs, the implementation of which in fitting procedures may supply FFs that come closer to true benchmarks for strong interaction theory.
Considering the predictions themselves, then by providing a unified set of parameter-free FFs for all reactions that contribute to $\pi, K$ production in hadron jets along with parton DFs for these same hadrons, the CSM FFs deliver a unique opportunity for developing a coherent reaction theory for high-energy processes \cite{Cui:2021mom}.
This could prove critical in making best use of data expected to be gathered at forefront and anticipated facilities.

Extensions of the present analyses to include proton FFs are underway, with a view to developing a comprehensive set of hadron FF predictions that is as encompassing as that which already exists for hadron DFs \cite{Cui:2020tdf, Lu:2022cjx, Lu:2023yna, Cheng:2023kmt, Yu:2024qsd}.
Heavy quark FFs are also being considered.

\begin{acknowledgements}
We are grateful to P.~Cheng, Z.-Q.\ Yao and W.-B.\ Yan for valuable discussions.
Work supported by:
National Natural Science Foundation of China (grant nos.\ 12135007, 12233002);
and Natural Science Foundation of Jiangsu Province (grant no.\ BK20220122).
\end{acknowledgements}

\begin{small}

\noindent\textbf{Data Availability Statement} Data will be made available on reasonable request.  [Authors' comment: All information necessary to reproduce the results described herein is contained in the material presented above.]
\medskip

\noindent\textbf{Code Availability Statement} Code/software will be made available
on reasonable request. [Authors' comment: No additional remarks.]

\end{small}


\end{document}